\documentclass[12pt]{article}
\usepackage[utf8]{inputenc}
\usepackage[T1]{fontenc}
\usepackage{hyperref}
\usepackage{url}
\usepackage{booktabs}
\usepackage{amsfonts}
\usepackage{nicefrac}
\usepackage{microtype}
\usepackage{lipsum}
\usepackage{graphicx}
\usepackage{natbib}
\usepackage{url} 
\usepackage{subfigure}
\usepackage{amsmath}
\graphicspath{ {./images/} }

\newtheorem{theorem}{Theorem}
\newtheorem{assumption}{Assumption}[section]

\title{Mediation Analysis for Sparse and Irregularly Spaced Longitudinal Outcomes with Application to the MrOS Sleep Study}

\author{
 Rui Ren$^\dag$, Haoyi Yang$^\ddag$, Qian Xiao$^\ast$, Lingzhou Xue$^\ddag$ and Yuan Huang$^\dag$\\
  $^\dag$Department of Biostatistics,  Yale University \\
  $^\ddag$Department of Statistics,   The Pennsylvania State University\\
  $^\ast$Department of Epidemiology, 
  University of Texas Health Sciences Center\\
}
\date{}
\begin{document}
\maketitle
\begin{abstract}
Mediation analysis has become a widely used method for identifying the pathways through which an independent variable influences a dependent variable via intermediate mediators. However, limited research addresses the case where mediators are high-dimensional and the outcome is represented by sparse, irregularly spaced longitudinal data. To address these challenges, we propose a mediation analysis approach for scalar exposures, high-dimensional mediators, and sparse longitudinal outcomes. This approach effectively identifies significant mediators by addressing two key issues: (i) the underlying correlation structure within the sparse and irregular cognitive measurements, and (ii) adjusting mediation effects to handle the high-dimensional set of candidate mediators. In the MrOS Sleep study, our primary objective is to explore lipid pathways that may mediate the relationship between rest-activity rhythms and longitudinal cognitive decline in older men. Our findings suggest a potential mechanism involving rest-activity rhythms, lipid metabolites, and cognitive decline, and highlight significant mediators identified through multiple testing procedures.
\end{abstract}

\noindent
{\it Keywords:} {Mediation analysis, high dimensional data, cognitive decline, rest-activity rhythms}

\renewcommand{\baselinestretch}{1.6}

\section{Introduction}
\label{sec: Introduction}
Mediation analysis aims to uncover the mechanisms by which an independent variable (e.g., exposure or treatment) affects a dependent variable (e.g., health outcome) through an intermediate variable, known as a mediator. Initially developed for scenarios involving a single mediator \citep{baron1986moderator}, mediation analysis has recently expanded significantly to accommodate a large number of mediators, driven by advances in data collection technologies. In high-dimensional settings, regularized approaches where only a few mediators are assumed to lie on the causal path between exposure and outcome, and multiple testing procedures have become prevalent \citep{blum2020challenges, zhao2022pathway, guo2022high}. For example, \cite{zhang2016estimating} and \citet{perera2022hima2} proposed methods for testing mediation effects in high-dimensional epigenetic studies through screening and regularized de-biasing procedures. Additionally, \citet{zhao2020sparse} introduced a sparse principal component-based approach to handle cases where high-dimensional mediators are causally dependent. Other advances include multiple-testing methods for high-dimensional mediation analysis \citep{dai2022multiple, liu2022large} and Bayesian inference techniques leveraging continuous shrinkage methods for high-dimensional data \citep{song2020bayesian, song2021bayesian}. Machine learning approaches have also been applied to mediation analysis to uncover more complex exposure-outcome relationships \citep{nath2023machine, wang2024dp2lm}. Diverse settings have prompted further innovation without the restricted scenario involving one exposure, multiple mediators, and one outcome. \citet{luo2020high} and \citet{zhang2021mediation} developed high-dimensional mediation analysis techniques for time-to-event outcome data using survival models. \citet{xu2023mediation} proposed a framework for mediation analysis where the mediator is represented as a graph. In cases where both the exposure and mediators are multivariate and potentially high-dimensional, \citet{zhao2022mediation} and \citet{zhao2022multimodal} introduced mediation analysis approaches for multiple exposures, multiple mediators, and continuous scalar outcomes, particularly in Alzheimer's disease research. Similarly, for settings involving multiple outcomes, \citet{zhao2023mediation} developed a multivariate-mediator and multivariate-outcome model to analyze multimodal neuroimaging data. 

There has been active research on longitudinal data in mediation analysis in recent years \citep{roth2013mediation, mittinty2020longitudinal}. \citet{bind2016causal} modeled the natural direct and indirect effects for different types of longitudinal mediators and outcomes, allowing interactions between the exposure and mediator. \citet{bind2017quantile} further introduced quantile mediation analysis to estimate the controlled direct and indirect effects of an exposure along percentiles of the mediator and outcome, addressing extreme values in medical endpoints with longitudinal mediators and outcomes. \citet{wei2024time} developed a time-varying mediation method to explore dynamic causal pathways in high-dimensional longitudinal DNA methylation (DNAm) data, identifying cytosine-phosphate-guanine sites where DNAm exhibits dynamic mediation effects. By modeling from a functional data analysis perspective, \citet{lindquist2012functional} and \citet{coffman2023causal} presented methods where the mediating variable is a continuous function rather than a single scalar, allowing for more detailed analysis of how mediators affect outcomes. \citet{zhao2018functional} extended this framework to cases where the exposure, mediator, and outcome are all continuous functions.  

An important remaining challenge for conducting mediation analysis in observational studies is that the longitudinal and repeated measurements of study outcomes are often collected sparsely with irregular time intervals \citep{bind2016causal, zeng2021causal}. For example, in the sub-cohort of the MrOS Sleep study (see more details in motivation data), the average number of repeated measurements for cognitive function over different follow-up periods per participant is 4.47, and the initial follow-up ages vary at enrollment, reflecting an irregularly spaced and extremely sparse data structure. This presents several new challenges for understanding the role of mediators in exposure-outcome associations: (i) the longitudinally measured outcomes may exhibit an unknown correlation structure, and (ii) understanding the mediation mechanism between scalar exposure and longitudinal outcomes via high-dimensional mediators. In this paper, we propose a mediation analysis method with high-dimensional mediators for longitudinal outcomes, suitable for both dense and sparse data, using a spline regression technique commonly used in functional data analysis.

The remainder of the paper is structured as follows: In Section \ref{sec: Motivation data}, we introduce the dataset that motivates this work. In Section \ref{sec: Mediation model setup} and \ref{sec: Estimation and inference}, we describe the mediation regression model and outline our estimation and inference procedures. Section \ref{sec: Numerical Study} demonstrates the performance of our method through simulation studies. In Section \ref{sec: Case Study}, we apply our method to examine the mediating effects of high-dimensional metabolic markers on the causal relationship between acrophase and cognitive function in the MrOS Sleep study. Finally, Section \ref{sec: Discussion} provides concluding remarks and suggests directions for future research.

\section{Motivation}
\label{sec: Motivation data}
Circadian rhythms play a crucial role in regulating numerous physiological processes and behaviors, and the rest-activity cycle is one of the most prominent behavioral manifestations of the circadian timing system. Over the past decade, a growing body of literature has linked impaired rest-activity rhythms (RAR) to various chronic diseases, including cognitive decline and dementia \citep{rogers2018rest, xiao2022nonparametric}. However, the potential mediation mechanisms remain largely underexplored \citep{xu2024association}. Identifying the underlying biological pathways that mediate the effects of impaired rest-activity rhythms on cognition will not only lead to a better understanding of disease etiology but may also enhance risk prediction and guide the development of preventive and therapeutic strategies.

The primary objective of this study is to examine the mediation relationship between rest-activity rhythms and cognitive decline via the lipid pathway in this population. Disorders in lipid metabolism, particularly cholesterol imbalances, have been linked to several neurodegenerative diseases, including Alzheimer’s, Parkinson’s, and Huntington’s disease, all of which are associated with cognitive decline symptoms \citep{block2010altered, liu2014lipid, alecu2019dysregulated}. Furthermore,  circadian rhythms are known to play a central role in lipid metabolism. For example, it has been shown that prolonged sleep restriction may modify cholesterol metabolism, and sleep disorders such as obstructive sleep apnea has been causally linked to the risk of dyslipidemia. Therefore, it is plausible to postulate that systematic alterations in lipid metabolism may mediate the effect of circadian disruption on cognition.

\begin{figure}[]
    \centering
    \includegraphics[width=0.55\linewidth]{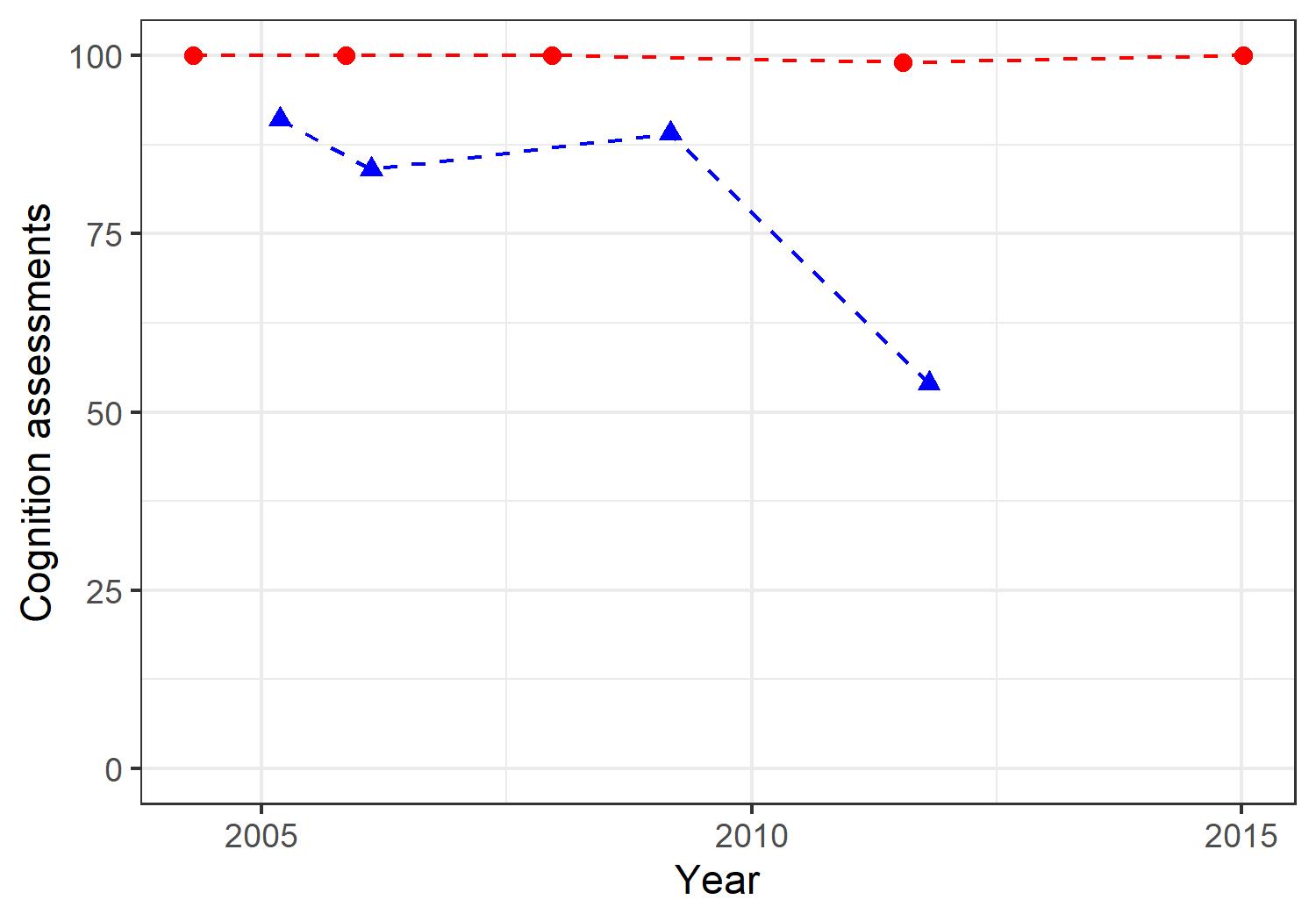}
    \caption{Longitudinal trajectories of cognition assessments for two randomly selected participants in the MrOS study.}
    \label{fig: Longitudinal trajectories for two randomly selected participants}
\end{figure}

Lipid metabolism is complex, involving hundreds or even thousands of metabolites and related pathways. Therefore, it is important to take a comprehensive approach to identify which lipid metabolites serve as significant mediators in the association between rest-activity rhythms and cognition. We took advantage of a large-scale metabolomics study in the MrOS Sleep, which measured 476 lipid markers from a wide range of pathways. Cognitive function was assessed using the modified Mini-Mental State Examination (3MS) at baseline and during follow-up evaluations. Rest-activity rhythms were measured by wrist actigraphy worn by participants for up to 5 consequtive 24-hour periods. We focused on acrophase (the time of daily peak activity) as the primary RAR variable. Cognitive assessments were conducted at baseline (approximately 2004-05) and during follow-up periods (approximately 2005-06, 2006-09, 2009-12, and 2012-16), with a maximum of five assessments per participant. The dataset includes 490 participants, of whom 260 completed four cognitive assessments and 230 completed five. The average number of repeated measurements per participant is 4.47. Figure \ref{fig: Longitudinal trajectories for two randomly selected participants} displays the longitudinal trajectories of cognitive assessments for two randomly selected participants. Cognitive function was assessed four or five times over a span of nearly 12 years, with significant variation in the time intervals between assessments across participants. 

Our proposed method addresses two primary challenges: (i) these features motivate the method to deal with irregular and sparse cognitive assessments comprehensively, offering a more complete understanding by accounting for the correlation structure among different assessments; and (ii) the mediation analysis model effectively identifies significant lipid metabolism mediators within the high-dimensional set of candidate mediators involved in the mediation pathway. 

\section{Mediation model setup}
\label{sec: Mediation model setup}
Mediation models are employed to evaluate the mechanisms through which an exposure influences an outcome. Let $X$ denote the exposure, $\boldsymbol{M}= (M_{1},\ldots,M_{p})^{\top}$ represent a vector of $p$ mediators, $\boldsymbol{Z}= (Z_{1},\ldots,Z_{q})^{\top}$ denote a vector of $q$ pre-exposure covariates, and consider a longitudinal outcome of interest. Suppose a sample of $n$ subjects is available. For each subject, observations of the longitudinal outcome are made at $n_{i}$ different time points, $\{t_{ij} \in [0,T], j=1,\ldots,n_{i}\}$, where $n_{i}$ can vary between subjects. A key to our method is modeling the observed outcome values $\boldsymbol{y}_{i} = (y_{i1}, \ldots, y_{i n_{i}} )^{\top}$ as realizations of a smooth underlying process $Y_{i}(t)$, $t \in [0,T]$, with normal measurement errors \citep{zeng2021causal}:
$$
y_{ij} = Y_{i}(t_{ij}) + \varepsilon_{ij}, \; \varepsilon_{ij} \sim N(0, \sigma_{\varepsilon_{ij}}^{2}).
$$
For each subject, the observation points are sparse and irregularly spaced across the time span. Additionally, denote the scalar covariates for each subject as $(x_{i}, \boldsymbol{m}_{i}^{\top}, \boldsymbol{z}_{i}^{\top})^{\top}$, where $\boldsymbol{m}_{i}= (m_{i1}, \ldots, m_{ip})^{\top}$, $\boldsymbol{z}_{i}= (z_{i1}, \ldots, z_{iq})^{\top}$, $i = 1, \ldots, n$. 

Instead of directly exploring the relationship between the exposure, mediators, and longitudinal outcomes $y_{ij}$'s, we investigate the relationship between the exposure, mediators, and stochastic process $Y_{i}(t)$. Adopting the counterfactual framework from causal inference literature \citep{pearl2000models, imbens2015causal}, let $\boldsymbol{M}_{x}$ represent the mediator values that would be observed if the exposure $X = x$ were set. Similarly, let $Y_{x}^{t}$ denote the potential outcome at time $t$ under $X = x$, and $Y_{x\boldsymbol{m}}^{t}$ denote the potential outcome under $X = x$ and $\boldsymbol{M} = \boldsymbol{m}$. For the outcome process until time $t$, we use bold notation: $\mathbf{Y}_{x}^{t} \equiv \{ Y_{x}^{s}, s \leq t \} \in \mathcal{R}^{[0,t]}$, and $\mathbf{Y}_{x\boldsymbol{m}}^{t} \equiv \{ Y_{x\boldsymbol{m}}^{s}, s \leq t \} \in \mathcal{R}^{[0,t]}$. We begin by stating two assumptions that are commonly invoked in the causal inference literature: the consistency and composition assumptions \citep{vanderweele2009conceptual}. For any $t \in [0, T]$, consistency implies that the potential outcome $Y_{x}^{t}$ is equal to the observed outcome $Y(t)$ when $X = x$ is observed. Similarly, $\boldsymbol{M}_{x}$ is equal to $\boldsymbol{M}$ when $X = x$ is observed. Additionally, the potential outcome $Y_{x\boldsymbol{m}}^{t}$ corresponds to the observed outcome $Y(t)$ under $X = x$ and $\boldsymbol{M} = \boldsymbol{m}$. On the other hand, composition assumes that the potential outcome $Y_{x}^{t}$ (intervening to set $X$ to $x$) is equal to the potential outcome $Y_{x\boldsymbol{m}}^{t}$ (intervening to set $X$ to $x$ and $M$ to the value it would have taken if $X$ had been $x$). That is, $Y_{x}^{t} = Y_{x\boldsymbol{m}}^{t}$, for any $t \in [0, T]$.
\begin{assumption}
    $\boldsymbol{M}_{x} \perp X\,|\,\boldsymbol{Z}$ for $\forall x$. That is, no unmeasured confounders between the exposure and the mediator.
    \label{assumption1}
\end{assumption}

\begin{assumption}
    $\mathbf{Y}_{x\boldsymbol{m}}^{t} \perp X\,|\,\boldsymbol{Z}$ for $\forall t, x, \boldsymbol{m}$. That is, no unmeasured confounders between the exposure and the outcome.
    \label{assumption2}
\end{assumption}

\begin{assumption}
    $\mathbf{Y}_{x\boldsymbol{m}}^{t} \perp \boldsymbol{M}\,|\,X, \boldsymbol{Z}$ for $\forall t, x, \boldsymbol{m}$. That is, no unmeasured confounders between the mediator and the outcome.
    \label{assumption3}
\end{assumption}

\begin{assumption}
    $\mathbf{Y}_{x\boldsymbol{m}}^{t} \perp \boldsymbol{M}_{x^{*}}\,|\,\boldsymbol{Z}$ for $\forall t, x, x^{*}$. That is, no exposure-induced confounding between the mediator and the outcome.
    \label{assumption4}
\end{assumption}

Under the above assumptions, the total effect (TE) on the outcome process at time $t$ is defined as:
$$
\begin{aligned}
    E [ Y_{x}^{t} - Y_{x^{*}}^{t}\,|\,\boldsymbol{z}] &= E [ Y_{x\boldsymbol{M}_{x}}^{t} - Y_{x^{*}\boldsymbol{M}_{x^{*}}}^{t}\,|\,\boldsymbol{z}] \\
    &= E [ Y_{x\boldsymbol{M}_{x}}^{t} - Y_{x\boldsymbol{M}_{x^{*}}}^{t}\,|\,\boldsymbol{z}] + E [ Y_{x\boldsymbol{M}_{x^{*}}}^{t} - Y_{x^{*}\boldsymbol{M}_{x^{*}}}^{t}\,|\,\boldsymbol{z}],
\end{aligned}
$$
where the first equality holds under the composition assumption. The term $E [ Y_{x\boldsymbol{M}_{x^{*}}}^{t} - Y_{x^{*}\boldsymbol{M}_{x^{*}}}^{t}\,|\,\boldsymbol{z}]$ is the natural direct (NDE), representing the expected difference in the outcome when changing the exposure value from $x$ to $x^{*}$, while fixing the mediator value at $\boldsymbol{M}_{x^{*}}$. The term $E [ Y_{x\boldsymbol{M}_{x}}^{t} - Y_{x\boldsymbol{M}_{x^{*}}}^{t}\,|\,\boldsymbol{z}]$ is the natural indirect effects (NIE), representing the expected difference in the outcome when changing the mediator value from $\boldsymbol{M}_{x}$ to $\boldsymbol{M}_{x^{*}}$, while holding the exposure value fixed at $x$. This decomposition remains valid even in the presence of nonlinearities and interactions.

We propose the following structural equations to model the mediation framework and evaluate time-varying mediation effects using time-varying coefficient regression models:
\begin{align}
M_{k} &= \kappa_{k} + \alpha_{k} X + \boldsymbol{\gamma}_{k}^{\top} \boldsymbol{Z} + \epsilon_{k}, \; k = 1, \ldots, p, \label{M_X}\\
Y(t) &= \mu(t) + \eta(t)X + \boldsymbol{\beta}(t)^{\top} \boldsymbol{M} + \boldsymbol{\theta}(t)^{\top} \boldsymbol{Z} + \varepsilon(t), \; \forall t \in [0,T], \label{Y_X+sumM}
\end{align}
where $\boldsymbol{\alpha} = (\alpha_{1}, \ldots, \alpha_{p})^{\top}$ and $\boldsymbol{\gamma}_{k} = (\gamma_{k1},\ldots,\gamma_{kq})^{\top}$ are regression coefficients in model (\ref{M_X}), while $\eta(t)$, $\boldsymbol{\beta}(t) = (\beta_{1}(t),\ldots,\beta_{p}(t))^{\top}$, and $\boldsymbol{\theta}(t) = (\theta_{1}(t), \ldots,\theta_{q}(t))^{\top}$ are time-varying regression coefficients in model (\ref{Y_X+sumM}). The terms $\kappa_{k}$ and $\mu(t)$ represent intercepts, and $\epsilon_{k}$ and $\varepsilon(t)$ are residual terms with zero mean. Under Assumptions \ref{assumption1}–\ref{assumption4}, and with correctly specified models (\ref{M_X}) and (\ref{Y_X+sumM}), the NDE and NIE can be estimated as:
\begin{equation}
\text{NDE} = (x - x^{*}) \cdot \eta(t), \quad
\text{NIE} = (x - x^{*}) \cdot \sum_{k=1}^{p} \alpha_{k}\beta_{k}(t).
\label{effect function}    
\end{equation}

\begin{figure}[]
    \centering
    \includegraphics[width=0.5\linewidth]{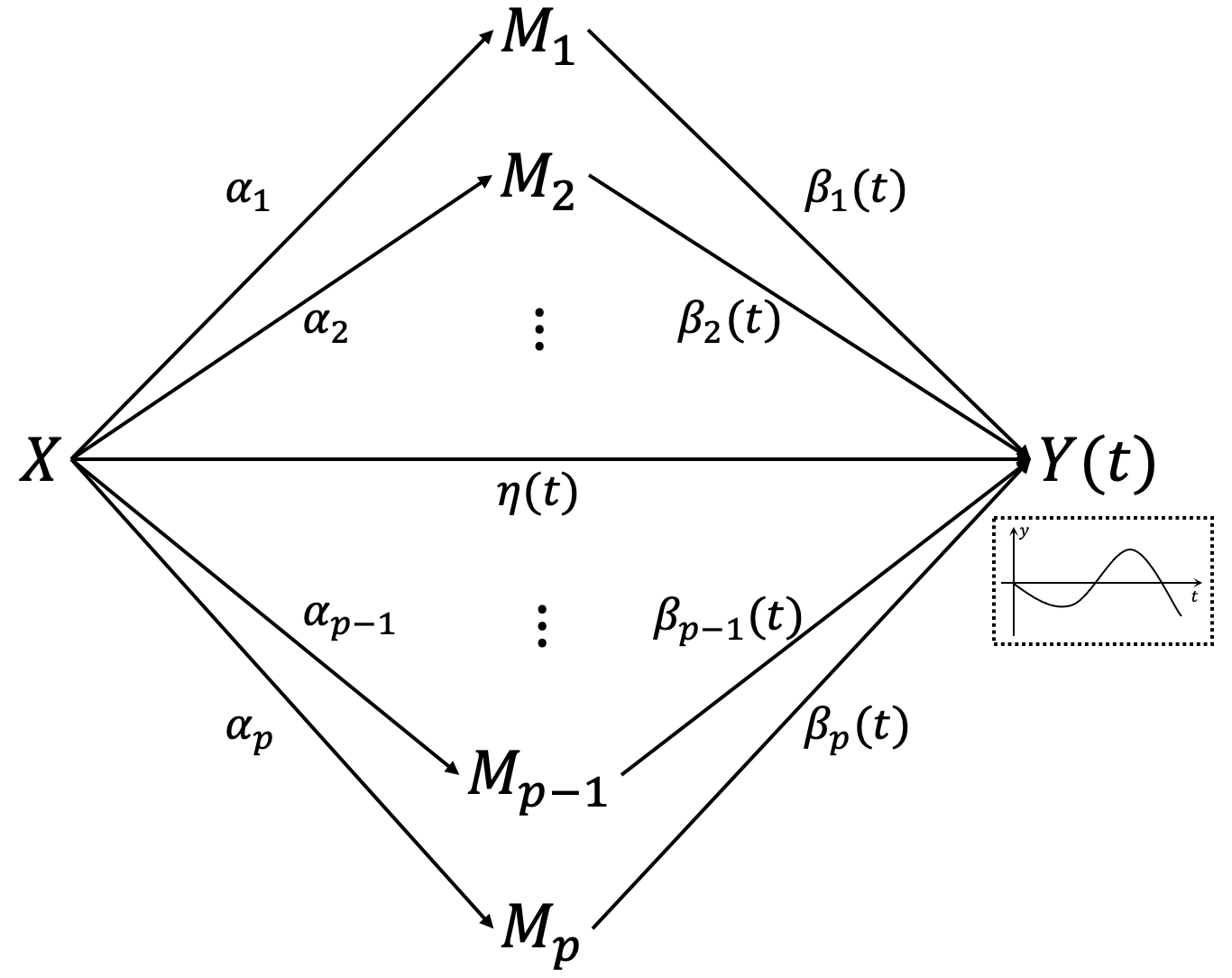}
    \caption{A scenario with high-dimensional mediators between exposure and outcome (confounding variables are omitted).}
    \label{fig: flowchart}
\end{figure}

\section{Estimation and inference}
\label{sec: Estimation and inference}
\subsection{Inference analysis with many potential mediators}
\label{sec: Inference analysis with many potential mediators}
In high-dimensional mediation analysis, where a large set of candidate mediators is considered, we perform inference analysis for triples $\{X, M_{k}, Y(t)\}$ to identify mediators that significantly mediate the effect of $X$ on $Y(t)$. This inference problem is formulated as the following multiple testing problem:
$$
H_{0}: \alpha_{k} = 0 \text{  or  } \beta_{k}(t) = 0 .
$$
The corresponding overall p-value for mediation is defined as:
\begin{equation}
    P_{k} = \max(P_{\alpha_{k}}, P_{\beta_{k}(t)} ).
    \label{joint pvalue}
\end{equation}
The multiple testing problem is decomposed into three disjoint components under the null hypothesis \citep{dai2022multiple, perera2022hima2}:
$$
\begin{aligned}
    &H_{00,k}: \, \alpha_{k} = 0 \text{  or  } \beta_{k}(t) = 0, \\
    &H_{01,k}: \, \alpha_{k} = 0 \text{  or  } \beta_{k}(t) \neq 0, \\
    &H_{10,k}: \, \alpha_{k} \neq 0 \text{  or  } \beta_{k}(t) = 0. \\
\end{aligned}
$$
That is, $P_{k}$ is a 3-component mixture distribution, for which \citet{dai2022multiple} proposed the following estimated False Discovery Rate (FDR) for testing mediation:
$$
\widehat{\text{FDR}}(\lambda) = \frac{\widehat{\pi}_{01} \lambda + \widehat{\pi}_{10} \lambda + \widehat{\pi}_{00} \lambda^{2} }{\max\{ 1, R(\lambda)\} /p},
$$
where $\widehat{\pi}_{01}$, $\widehat{\pi}_{10}$, and $\widehat{\pi}_{00}$ are the estimates of proportions $H_{01,k}$, $H_{10,k}$ and $H_{00,k}$, respectively, and $R(\lambda) = V_{00}(\lambda) + V_{01}(\lambda) + V_{10}(\lambda) + V_{11}(\lambda)$ represents the number of rejections under threshold $\lambda$, where $V_{00}(\lambda) = \# \{ P_{k} \leq \lambda | H_{00} \}$, $V_{01}(\lambda) = \# \{ P_{k} \leq \lambda | H_{01} \}$, $V_{10}(\lambda) = \# \{ P_{k} \leq \lambda | H_{10} \}$, $V_{11}(\lambda) = \# \{ P_{k} \leq \lambda | H_{11} \}$ for $\lambda \in [0,1]$. We define the significant threshold for $P_{k}$ as $\widehat{\lambda}_{b} = \sup \big\{ \lambda: \widehat{\text{FDR}}(\lambda) \leq \lambda_{b} \big\}$, to control the FDR at level $b$. Then, $\widehat{\mathcal{I}} = \big\{ k:P_{k} \leq \widehat{\lambda}_{b}, \, k=1,\ldots, p\big\}$ provides the estimated index set of significant mediators.

To obtain the overall p-value for each mediator ($k$ in $1, \ldots, p$), we need both $P_{\alpha_{k}}$ and $P_{\beta_{k}(t)}$. Here $P_{\alpha_{k}} = 2\left\{1-\Phi\big( | \widehat{\alpha}_{k} | / \widehat{\sigma}_{\alpha_{k}}\big) \right\}$, with $\widehat{\alpha}_{k}$ and $\widehat{\sigma}_{\alpha_{k}}$ denoting the ordinary least squares (OLS) estimators obtained from equation (\ref{M_X}). Regarding $P_{\beta_{k}(t)}$, which concerns hypothesis testing for the time-varying effect $\beta_{k}(t)$, we propose a permutation test specifically designed for time-varying coefficient models to compute $P_{\beta_{k}(t)}$. Consider the following marginal time-varying regression model with the $k$-th candidate mediator:
\begin{equation}
    y_{i}(t_{ij}) = \mu_{k}^{*}(t_{ij}) + \eta_{k}^{*}(t_{ij})x_{i} + \beta_{k}^{*}(t_{ij}) m_{ik} + \boldsymbol{\theta}_{k}^{*}(t_{ij})^{\top} \boldsymbol{z}_{i} + \varepsilon_{ki}^{*}(t_{ij}),
\label{Y_X+M_descrete}
\end{equation}
where $i = 1,\ldots, n$ and $j = 1, \ldots, n_{i}$. In functional data analysis, the Karhunen-Loève (KL) expansion is commonly employed, allowing the time-varying coefficient functions $\mu_{k}^{*}(t)$, $\eta_{k}^{*}(t)$, $\beta_{k}^{*}(t)$, and $\{\theta_{kl}^{*}(t)\}_{l=1,\ldots,q}$ to be expressed as:
$$
\begin{aligned}
\mu_{k}^{*}(t) = \sum_{r=1}^{\infty} \xi_{\mu_{k}, r \,} \psi_{\mu_{k}, r}(t),& \quad
\eta_{k}^{*}(t) = \sum_{r=1}^{\infty} \xi_{\eta_{k}, r \,} \psi_{\eta_{k}, r}(t), \\
\beta_{k}^{*}(t) = \sum_{r=1}^{\infty} \xi_{\beta_{k}, r \,} \psi_{\beta_{k}, r}(t),& \quad
\theta_{kl}^{*}(t) = \sum_{r=1}^{\infty} \xi_{\theta_{kl}, r \,} \psi_{\theta_{kl}, r}(t),
\end{aligned}
$$
where $\{\psi_{\cdot, r}\}_{r=1,\ldots, \infty}$ is a set of pre-specified orthogonal basis functions (e.g., B-splines, Fourier basis functions, Wavelet basis functions), and $\xi_{\cdot, r}$'s are the corresponding basis coefficients. In practice, the truncated KL expansion is employed, yielding the following spline regression model:
\begin{align}
    y_{i}(t_{ij}) = & \sum_{r=1}^{L_{\mu_{k}}} \xi_{\mu_{k}, r \,} \psi_{\mu_{k}, r}(t_{ij}) + \sum_{r=1}^{L_{\eta_{k}}} x_{i} \, \xi_{\eta_{k}, r \,} \psi_{\eta_{k}, r}(t_{ij}) \notag \\ 
    +& \sum_{r=1}^{L_{\beta_{k}}} m_{ik} \, \xi_{\beta_{k}, r \,} \psi_{\beta_{k}, r}(t_{ij}) 
    + \sum_{l=1}^{q} \sum_{r=1}^{L_{\theta_{k}}} z_{il} \, \xi_{\theta_{kl}, r \,} \psi_{\theta_{kl}, r}(t_{ij}) + \varepsilon_{ki}^{*}(t_{ij}),
    \label{Y_X+M_varying}
\end{align}
where $L_{\cdot}$ denotes the number of basis functions. Note that the basis functions for different coefficient functions can vary, as can the number of basis functions. The error term $\varepsilon_{ki}^{*}(t_{ij})$ is allowed to have a time-varying variance $V(t)$ and to be correlated across $t_{ij}$ for the same subject, providing flexibility in model construction. This within-subject correlation is characterized by an $n_{i} \times n_{i}$ correlation matrix $\mathbf{R}_{i}(\rho)$, where $\rho$ is a parameter that fully defines the correlation structure—such as the commonly used diagonal (independent), uniform, or autoregressive (AR) correlation structures \citep{huang2004polynomial}. 

To account for both the time-varying variance and within-subject correlation, we specify the error covariance matrix as 
$$\mathbf{C}_{i} = \mathbf{V}_{i}^{1/2} \mathbf{R}_{i}(\rho) \mathbf{V}_{i}^{1/2},$$ where $\mathbf{V}_{i}$ is a diagonal matrix containing $V(t_{ij})$ on its diagonal. Furthermore, we adopt the weighted least squares (WLS) approach with a weight matrix $\mathbf{W}_{i} = n_{i}^{-1} \mathbf{C}_{i}^{-1}$ to improve estimation efficiency in model (\ref{Y_X+M_varying}). The construction of this weight matrix follows procedures similar to those proposed in \citet{chu2016feature}. Specifically, we first apply the ordinary least squares method to the spline regression model under the null hypothesis $\beta_{k}^{*}(t) = 0$ in equation (\ref{Y_X+M_descrete}) to obtain residuals $\hat{\varepsilon}_{i}(t_{ij})$, which are then used to estimate the time-varying variance $\{\widehat{V}(t_{i1}), \ldots, \widehat{V}(t_{in_{i}})\}$. The correlation parameter $\widehat{\rho}$ is subsequently estimated based on these residuals, with detailed calculation procedures provided in Section S2 of the supplementary file. 

Let $\boldsymbol{\psi}_{\cdot} = (\psi_{\cdot,1}(t), \ldots, \psi_{\cdot, L_{\cdot}}(t))$ denote the vector of spline basis functions, and define the corresponding spline matrix as:
$$
\boldsymbol{\Psi}_{ki}(t) = \left(\begin{array}{cccccc}
\boldsymbol{\psi}_{\mu_{k}}(t) & 0 & 0 & 0 & \ldots & 0 \\ 
0 & \boldsymbol{\psi}_{\eta_{k}}(t) & 0 & 0 & \ldots & 0 \\ 
0 & 0 & \boldsymbol{\psi}_{\beta_{k}}(t) & 0 & \ldots & 0 \\ 
0 & 0 & 0 & \boldsymbol{\psi}_{\theta_{k1}}(t) & \ldots & 0 \\ 
\vdots & \vdots & \vdots & \vdots & \ddots & \vdots \\ 
0 & 0 & 0 & 0 & \ldots & \boldsymbol{\psi}_{\theta_{kq}}(t)
\end{array}\right).
$$
To express the estimators of model (\ref{Y_X+M_varying}) in a matrix form, let $\mathbf{D}_{kij}^{\top} = (1, x_{i}, m_{ik}, \boldsymbol{z}_{i}^{\top})\boldsymbol{\Psi}_{ki}(t_{ij})$, and define $\mathbf{D}_{ki} = (\mathbf{D}_{ki1},\ldots, \mathbf{D}_{kin_{i}})^{\top}$. Further, denote the spline coefficients as follows: $\boldsymbol{\xi}_{\mu_{k}} = (\xi_{\mu_{k},1}, \ldots, \xi_{\mu_{k}, L_{\mu_{k}}})^{\top}$, $\boldsymbol{\xi}_{\eta_{k}} = (\xi_{\eta_{k},1}, \ldots, \xi_{\eta_{k}, L_{\eta_{k}}})^{\top}$, $\boldsymbol{\xi}_{\beta_{k}} = (\xi_{\beta_{k},1}, \ldots, \xi_{\beta_{k}, L_{\beta_{k}}})^{\top}$ for the mediator $M_{k}$, and $\boldsymbol{\xi}_{\theta_{kl}} = (\xi_{\theta_{kl},1}, \ldots, \xi_{\theta_{kl}, L_{\theta_{k}}})^{\top}$ for the $l$-th covariate $Z$-variable, where $l = 1, \ldots, q$. Using the estimated weight matrix $\widehat{\mathbf{W}}_{i}$, we apply the WLS method to estimate the vector of regression coefficients $\boldsymbol{\xi}_{k} = (\boldsymbol{\xi}_{\mu_{k}}^{\top}, \boldsymbol{\xi}_{\eta_{k}}^{\top}, \boldsymbol{\xi}_{\beta_{k}}^{\top}, \boldsymbol{\xi}_{\theta_{k1}}^{\top}, \ldots, \boldsymbol{\xi}_{\theta_{kq}}^{\top} )^{\top}$ in model (\ref{Y_X+M_varying}) as:
$$
\widehat{\boldsymbol{\xi}}_{k} = \left( \frac{1}{n} \sum_{i=1}^{n} \mathbf{D}_{ki}^{\top} \widehat{\mathbf{W}}_{i} \mathbf{D}_{ki} \right)^{-1} \left( \frac{1}{n} \sum_{i=1}^{n} \mathbf{D}_{ki}^{\top} \widehat{\mathbf{W}}_{i} \boldsymbol{y}_{i} \right) .
$$
The estimated $\boldsymbol{y}_{i}$ can be obtained by $\widehat{\boldsymbol{y}}_{i}^{(k)} = \mathbf{D}_{ki} \widehat{\boldsymbol{\xi}}_{k}$, allowing us to calculate the weighted residual sum of squares (WRSS) as:
$$
\widehat{\text{WRSS}}_{k} = \sum_{i=1}^{n} \left( \boldsymbol{y}_{i} - \widehat{\boldsymbol{y}}_{i}^{(k)} \right)^{\top} \widehat{\mathbf{W}}_{i} \left( \boldsymbol{y}_{i} - \widehat{\boldsymbol{y}}_{i}^{(k)} \right).
$$
For this study, we employ a B-spline expansion for each coefficient function and, for simplicity, assume the same number of basis functions across all cases, denoted as $L_{n}$.

The smaller the value of the weighted residual sum of squares, the stronger the marginal association between the $k$-th mediator and the response. To test whether there is any association between the mediator and the functional response $y_{i}(t)$, we can formulate it as testing the null hypothesis, $H_{k0}:\, \boldsymbol{\xi}_{\beta_{k}} = \mathbf{0}$, against the alternative hypothesis $H_{k1}: \, \boldsymbol{\xi}_{\beta_{k}} \neq \mathbf{0}$. We define $\text{WRSS}_{0}$ and $\text{WRSS}_{k}$ as the weighted residual sum of squares of the model without and with the $k$-th candidate mediator, respectively. Specifically, $\text{WRSS}_{0}$ is calculated under $H_{k0}$, with calculation details similar to those of $\text{WRSS}_{k}$ except that the $M_{k}$-variable is excluded from this model. The test statistic can be defined as:
\begin{equation}
    F_{k} = \frac{ \text{WRSS}_{0} - \text{WRSS}_{k} }{\text{WRSS}_{k}}.
    \label{F-statistics}
\end{equation}
We employ a resampling approximation permutation test to evaluate the importance of the $M_{k}$ variable using the F-statistic defined in (\ref{F-statistics}). By randomly generating many datasets in which the $k$-th mediator variable values are permuted, we obtain the test statistic $F_{k}^{(s)}$ for the $s$-th replication, where $s = 1, \ldots, S$. The p-value $P_{\beta_{k}(t)}$ is estimated as the proportion of times the test statistic is as extreme as the one observed, based on simulating the test statistic a large number of times 
$S$ under the null model, that is, $\frac{\sum_{s=1}^{S} I(F_{k}^{(s)} > F_{k}) }{S}$. Note that the estimation depends on the choice of the correlation structure $\mathbf{R}_{i}(\rho)$; we have investigated the impact of the choice through simulations in Section \ref{sec: Numerical Study}. 

We define the estimated false discovery rate $\widehat{\text{FDR}}({\lambda})$ using the estimated proportions of component null hypotheses as follows:
$$
\widehat{\text{FDR}}({\lambda})= \frac{\hat{\pi}_{00}\lambda^2 + \hat{\pi}_{01}\lambda + \hat{\pi}_{10}\lambda}{(R(t) \vee 1)/p}.
$$
where $\hat{\pi}_{00}$, $\hat{\pi}_{01}$, $\hat{\pi}_{10}$ are the estimated proportions of the joint null components, and $R(\lambda)$ denotes the number of rejections under threshold $\lambda$. After obtaining $P_{\alpha_{k}}$'s and $P_{\beta_{k}(t)}$'s, we can calculate $\widehat{\pi}_{01}$, $\widehat{\pi}_{10}$, $\widehat{\pi}_{00}$ and $\widehat{\lambda}_{b}$ using the R package \texttt{HDMT} \citep{dai2022multiple}. The set of significant mediators is then identified as $\widehat{\mathcal{I}} = \big\{ k:P_{k} \leq \widehat{\lambda}_{b}, \, k=1,\ldots, p\big\}$ under the FDR control at level $b$. 

Theorem \ref{theorem1} shows that we have the desired asymptotic control property for the estimated false discovery rate $\widehat{\text{FDR}}({\lambda})$. The proof of Theorem \ref{theorem1} is provided in Section S1 of the supplementary file due to space constraints. 

\begin{theorem}
Under the correctly specified time-varying model, assume that $P_{\alpha_k}$ and $P_{\beta_k(t)}$ are asymptotically independent for each hypothesis $k$. For sufficiently large permutation times $S$ and when $p$ is bounded by a polynomial function of $L_n$, under Assumptions S1.1-S1.8 in supplementary file, for a threshold $\lambda_b$ chosen such that $\widehat{\text{FDR}}(\lambda) \leq b$, where $b$ is a pre-specified target FDR level, we have the asymptotic control property:
\[
\limsup_{p \to \infty} \widehat{\text{FDR}}({\lambda}) \leq b.
\]
\label{theorem1}
\end{theorem}

\subsection{Mediation analysis with screened significant mediators}
\label{sec: Mediation analysis with screened significant mediators}
Suppose $p_{0}$ mediators are identified as having significant mediation effects. We then rewrite the mediation framework as:
\begin{align}
M_{k} &= \kappa_{k} + \alpha_{k} X + \boldsymbol{\gamma}_{k}^{\top} \boldsymbol{Z} + \epsilon_{k}, \; k = 1, \ldots, p_{0}, \label{M_X_screened}\\
Y(t) &= \mu(t) + \eta(t)X + \boldsymbol{\beta}(t)^{\top} \boldsymbol{M} + \boldsymbol{\theta}(t)^{\top} \boldsymbol{Z} + \varepsilon(t), \; \forall t \in [0,T]. \label{Y_X+sumM_screened}
\end{align}

As in the previous section, we use a truncated KL expansion to rewrite equation (\ref{Y_X+sumM_screened}) as:
$$
y_{i}(t_{ij}) = \sum_{r=1}^{L_{n}} \zeta_{\mu, r \,} \psi_{r}(t_{ij}) + \sum_{r=1}^{L_{n}} x_{i} \, \zeta_{\eta, r \,} \psi_{r}(t_{ij}) + \sum_{k=1}^{p_{0}} \sum_{r=1}^{L_{n}} m_{ik} \, \zeta_{\beta_{k}, r \,} \psi_{r}(t_{ij}) + \sum_{l=1}^{q} \sum_{r=1}^{L_{n}} z_{il} \, \zeta_{\theta_{l}, r \,} \psi_{r}(t_{ij}) + \varepsilon_{i}(t_{ij}),
$$
where $L_{n}$ is the truncated number of basis functions, $\{\psi_{r}\}_{r=1,\ldots, L_{n}}$ denotes the orthonormal basis set, and $\zeta_{\cdot, r}$ are the corresponding basis coefficients. Let $\boldsymbol{\psi} = (\psi_{1}(t), \ldots, \psi_{L_{n}}(t))$, and define the spline matrix $\boldsymbol{\Psi}(t)$ with $(p_{0}+q+2)L_{n}$ rows as:
$$
\boldsymbol{\Psi}(t) = \left(\begin{array}{cccc}
\boldsymbol{\psi}(t) & 0 & \ldots & 0 \\ 
0 & \boldsymbol{\psi}(t) & \ldots & 0 \\ 
\vdots & \vdots & \ddots & \vdots \\ 
0 & 0 & \ldots & \boldsymbol{\psi}(t)
\end{array}\right).
$$
Define $\mathbf{D}_{ij}^{\top} = (1, x_{i}, m_{i1},\ldots,m_{ip_{_0}},\boldsymbol{z}_{i}^{\top})\boldsymbol{\Psi}(t_{ij})$, and $\mathbf{D}_{i} = (\mathbf{D}_{i1},\ldots, \mathbf{D}_{in_{i}})^{\top}$. Denote the spline coefficients as $\boldsymbol{\zeta}_{\mu} = (\zeta_{\mu,1}, \ldots, \zeta_{\mu, L_{n}})^{\top}$, $\boldsymbol{\zeta}_{\eta} = (\zeta_{\eta,1}, \ldots, \zeta_{\eta, L_{n}})^{\top}$, $\boldsymbol{\zeta}_{\beta_{k}} = (\zeta_{\beta_{k},1}, \ldots, \zeta_{\beta_{k}, L_{n}})^{\top}$ for the mediator $M_{k}$, and $\boldsymbol{\zeta}_{\theta_{l}} = (\zeta_{\theta_{l},1}, \ldots, \zeta_{\theta_{l}, L_{n}})^{\top}$ for the $l$-th covariate $Z$-variable. 

We use a weighted least squares estimator, with the weight matrix $\widehat{\mathbf{W}}_{i}$ described in Section \ref{sec: Inference analysis with many potential mediators}, to estimate the regression coefficients 
$$
\boldsymbol{\zeta} = (\boldsymbol{\zeta}_{\mu}^{\top}, \boldsymbol{\zeta}_{\eta}^{\top}, \boldsymbol{\zeta}_{\beta_{1}}^{\top}, \ldots, \boldsymbol{\zeta}_{\beta_{p_{_0}}}^{\top}, \boldsymbol{\zeta}_{\theta_{1}}^{\top}, \ldots, \boldsymbol{\zeta}_{\theta_{q}}^{\top} )^{\top} ,
$$
via
$$
\widehat{\boldsymbol{\zeta}} = \left( \frac{1}{n} \sum_{i=1}^{n} \mathbf{D}_{i}^{\top} \widehat{\mathbf{W}}_{i} \mathbf{D}_{i} \right)^{-1} \left( \frac{1}{n} \sum_{i=1}^{n} \mathbf{D}_{i}^{\top} \widehat{\mathbf{W}}_{i} \boldsymbol{y}_{i} \right) .
$$
These estimated coefficients are then used for the final estimation. Recalling the definitions from equation~\eqref{effect function}, the natural direct and indirect effects are estimated as:
\begin{equation}
\widehat{\text{NDE}} = (x - x^{*}) \cdot  \boldsymbol{\psi}(t) \widehat{\boldsymbol{\zeta}}_{\eta}, \quad
\widehat{\text{NID}} = (x - x^{*}) \cdot \sum_{k=1}^{p_{0}} \widehat{\alpha}_{k}\boldsymbol{\psi}(t) \widehat{\boldsymbol{\zeta}}_{\beta_{k}},
\label{effect estimators}
\end{equation}
where $\widehat{\alpha}_{k}$ is the OLS estimator equation~\eqref{M_X_screened}.

\section{Numerical study}
\label{sec: Numerical Study}
\subsection{Design of study}
\label{subsec: Design of Study}
We designed a simulation study to assess the performance of our proposed method. Consider the mediation model introduced in Section \ref{sec: Mediation model setup},
$$
\begin{aligned}
M_{k} &= \kappa_{k} + \alpha_{k} X + \boldsymbol{\gamma}_{k}^{\top} \boldsymbol{Z} + \epsilon_{k}, \; k = 1, \ldots, p, \\
Y(t) &= \mu(t) + \eta(t)X + \boldsymbol{\beta}(t)^{\top} \boldsymbol{M} + \boldsymbol{\theta}(t)^{\top} \boldsymbol{Z} + \varepsilon(t), \; \forall t \in [0,T] .
\end{aligned}
$$

We considered a sample size $n=100$ and $p=50$ candidate mediators. We generated the exposure $X$ from $N(0,2)$. For the confounding covariates $\boldsymbol{Z} = (Z_{1}, Z_{2})^{\top}$, both $Z_{1}$ and $Z_{2}$ were independently generated from $N(0,2)$. We set $\kappa_{k} = 0$ and $\boldsymbol{\gamma}_{k} = (0.3, 0.3)^{\top}$ for each $k = 1, \ldots, p$; and let $\mu(t) = 0$, $\theta_{1}(t) = 0.5\cos(3 \pi t)$, $\theta_{2}(t) = 0.4\sin(3 \pi t)$, $\eta(t) = 0.3\exp(0.5 t^{2})$, $\alpha_{1} = 0.35$, $\alpha_{2} = 0.4$, $\alpha_{3} = 0.25$, $\alpha_{4} = 0.3$, $\alpha_{5} = 0.15$ and $\alpha_{k} = 0$ for all other $k$'s; $\beta_{1}(t) = 0.5 \sin(\pi t)$, $\beta_{2}(t) = -0.4 \sin(2 \pi t)$, $\beta_{3}(t) = - 0.4 \cos(2 \pi t)$, $\beta_{4}(t) = 0.5(1.2 - t)$, $\beta_{6}(t) = 0.5 \delta \cos(\pi t)\mathbf{I}_{\{t \geq 0.5\} }$, and $\beta_{k}(t) = 0$ for all other $k$'s, where $\mathbf{I}_{\{\cdot\}}$ represents the indicator function. From these definitions, we have: (1) $\alpha_{k} \beta_{k}(t) \neq 0$ for $k = 1, \ldots, 4$; (2) $\alpha_{k} \neq 0$ but $\beta_{k}(t) = 0$ for $k = 5$; (3) $\alpha_{k} = 0$ but $\beta_{k}(t) \neq 0$ for $k = 6$; and (4) $\alpha_{k} = 0$ and $\beta_{k}(t) = 0$ for $k > 6$. That is, $p_{0} = 4$ mediators were set to have significant mediation effects. The mediator error terms $\boldsymbol{\epsilon} = (\epsilon_{1}, \ldots, \epsilon_{p})^{\top}$ were drawn from a multivariate normal distribution $N(\mathbf{0}, \Sigma_{\epsilon})$, where $\Sigma_{\epsilon} = \big( 0.1^{|k_{1} - k_{2}|} \big)_{k_{1}, k_{2}}$ with $k_{1}$ and $k_{2}$ indicating the row and column indices. The outcome error terms $\varepsilon_{i}(t_{ij})$ were generated from a zero mean Gaussian process with varying correlation structures across the following four cases, where $(\rho_{1}, \rho_{2}) = (0.8, 0.3)$, and for $j, j_{1}, j_{2} \in \{ 1, \ldots, n_{i} \}$ and $j_{1} \neq j_{2}$:
\begin{itemize}
    \item [Case 1]
    (diagonally independent) $$\text{Var}\big(\varepsilon_{i}(t_{ij})\big) = 2, \quad \text{cor}\big( \varepsilon_{i}(t_{ij_{1}}), \varepsilon_{i}(t_{ij_{2}}) \big) = 0;$$

    \item [Case 2]
    (autoregressive of order 1) $$\text{Var}\big(\varepsilon_{i}(t_{ij})\big) = 2, \quad \text{cor}\big( \varepsilon_{i}(t_{ij_{1}}), \varepsilon_{i}(t_{ij_{2}}) \big) = \rho_{1}^{|j_{1} - j_{2}|};$$

    \item [Case 3]
    (uniform) $$\text{Var} \big( \varepsilon_{i}(t_{ij}) \big) = 2, \quad \text{cor}\big( \varepsilon_{i}(t_{ij_{1}}), \varepsilon_{i}(t_{ij_{2}}) \big) = \rho_{2};$$

    \item [Case 4]
    (compound symmetry) $$\text{Var} \big( \varepsilon_{i}(t_{ij}) \big) = 1.5 + 2t_{ij}^{2}, \quad \text{cor}\big( \varepsilon_{i}(t_{ij_{1}}), \varepsilon_{i}(t_{ij_{2}}) \big) = 0.5 \rho_{1}^{|j_{1} - j_{2}|} + 0.5\rho_{2}.$$
\end{itemize}

Furthermore, for each case, we considered three scenarios for different types of sampling for time: (i) Dense measurements (Scenario 1), where $t_{ij}$ values were uniformly sampled from $[0,1]$ with $n_{i}=100$ for each $i = 1,\ldots,n$; (ii) Sparse and irregularly spaced measurements (Scenario 2), where $t_{ij}$ values were uniformly sampled from $[0,1]$ with $n_{i}=10$; (iii) Sparse and irregularly spaced measurements with varying numbers of observations (Scenario 3), where 10\% of subjects had $\{t_{ij}\}_{j=1,\ldots,n_{i}}$ uniformly sampled from $[0,0.7]$ with $n_{i}=7$, 20\% from $[0,0.8]$ with $n_{i}=8$, 30\% from $[0,0.9]$ with $n_{i}=9$, and 40\% from $[0,1]$ with $n_{i}=10$. 

We conducted $G=100$ simulation replications and performed $S = 1000$ permutations for each testing procedure. 

\subsection{Implementation and the alternative}
Our proposed method involves two key components: the correlation structure of the error terms $\varepsilon(t)$, and the number of basis functions used in the spline regression. 

For the correlation matrix $\mathbf{R}_{i}(\rho)$, we considered three commonly used structures: diagonal independent, uniform, and autoregressive of order 1. Depending on the selected correlation structure, our proposed method takes one of the following forms: Diagonal, Uniform, or AR. The AR structure captures the decreasing correlation between repeated measurements as the time interval increases, making it most appropriate for equally spaced measurements. To better accommodate unequally spaced repeated measurements that are more aligned with the data generation mechanism, we also consider a power correlation structure, where $d_{j_1 j_2}$ denotes the Euclidean distance between the $j_1$-th and $j_2$-th measurement time points. This results in an additional variant of our method, referred to as ``Proposed (Power)," which is based on the power correlation structure. The specifications and detailed calculations for these four correlation structures are provided in Section S2 of the supplementary file.

For the number of basis functions $L_{n}$ used in the spline regression, a larger value allows for more accurate estimation of the varying coefficients but comes at the cost of increased variance. The selection of $L_{n}$ involves three tuning parameters: the degree of the splines $\vartheta$, the number of interior knots $\varpi_{n}$, and the positions of the interior knots. Across all simulation setups, we set the spline degree to three (i.e., cubic splines), which is the most commonly used option. Since the time points are transformed to be approximately uniform over $[0, 1]$, we use equally spaced knots. To select the number of interior knots, we employ 10-fold cross-validation over the candidate set $\varpi_{n} \in \{ 1,2,\cdots,8\}$, and determine the optimal number of basis functions as $L_{n} = \varpi_{n} + \vartheta + 1$ by minimizing the prediction error. Our results confirm that choosing $L_{n} \in \{5, 6, \ldots, 12\}$ is sufficient for accurate model fitting. Further details are provided in Section S3 of the supplementary file.

Throughout the simulation, we considered an alternative approach, the FoSR method, which uses a pointwise F-like statistic from the FoSR test introduced by \citet{reiss2010fast} and \citet{reiss2011extracting}, combined with the same multiple-testing procedure to identify significant mediators. 

All simulations were conducted in R (version 4.2.0), primarily using the fda package for functional data analysis and the HDMT package for multiple testing procedures. 

\subsection{The performance of the proposed permutation test}
\label{subsec: The Performance of Proposed Permutation Test}

\begin{figure}[]
	\centering
	\subfigure{
          \rotatebox{90}{\scriptsize{~~~~~~~~~~~~~~Case 1}}
	\begin{minipage}[t]{0.3\textwidth}
		\includegraphics[width=\textwidth]{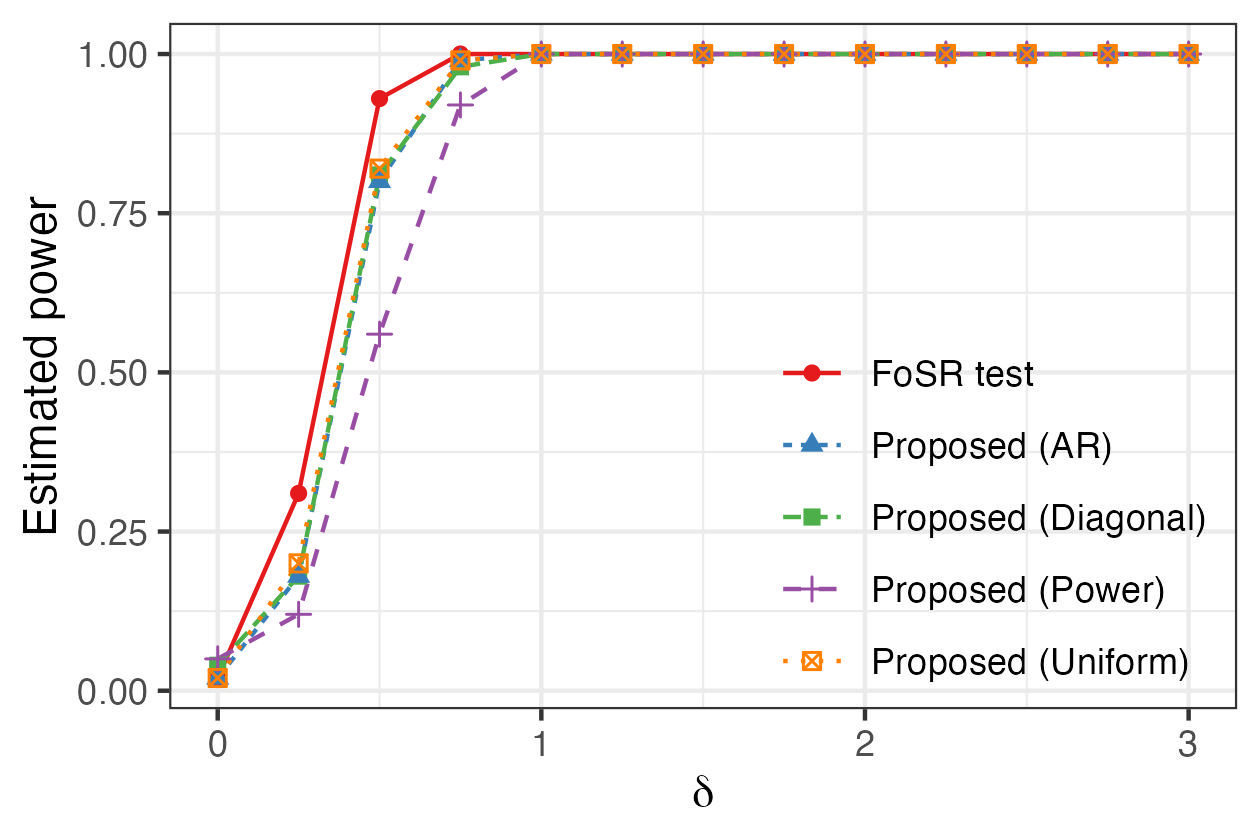}
	\end{minipage}
	}
	\subfigure{
        \begin{minipage}[t]{0.3\textwidth}
		\includegraphics[width=\textwidth]{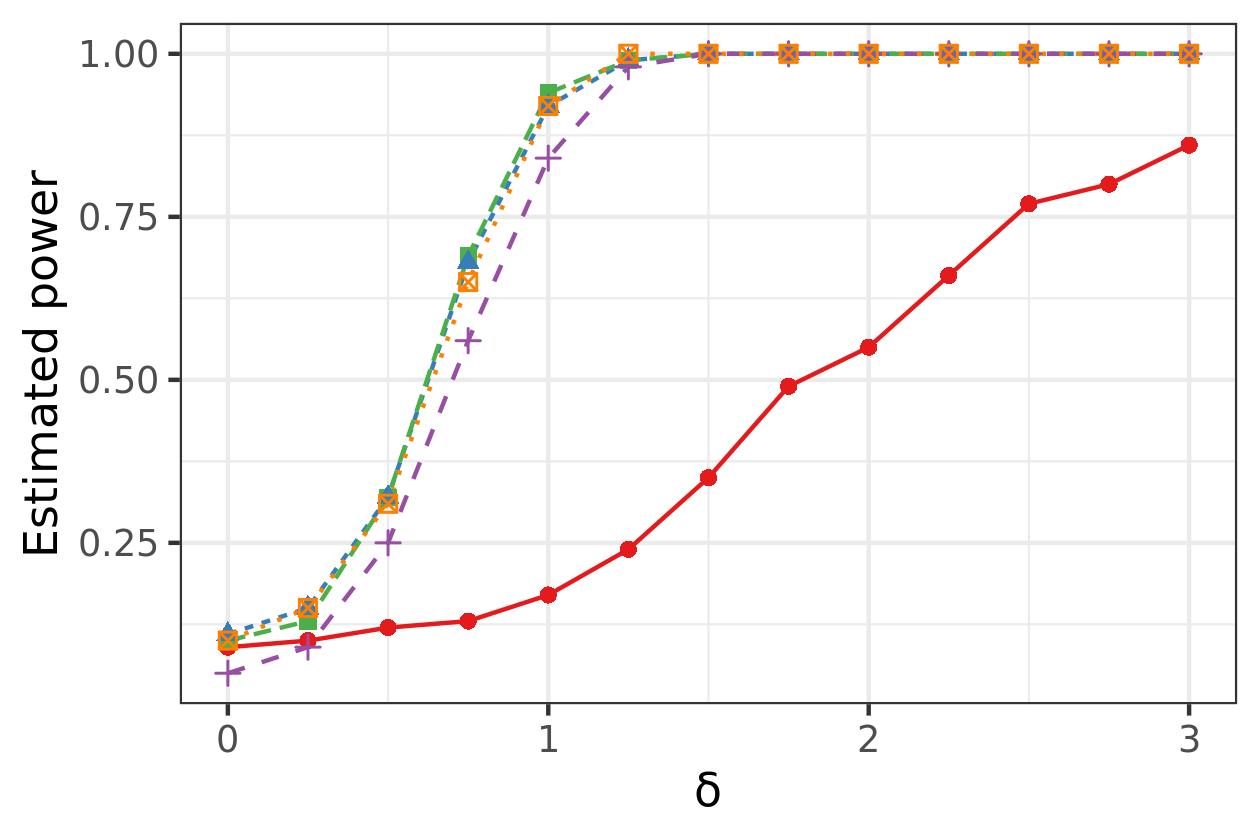}
	\end{minipage}
        }
        \subfigure{
        \begin{minipage}[t]{0.3\textwidth}
		\includegraphics[width=\textwidth]{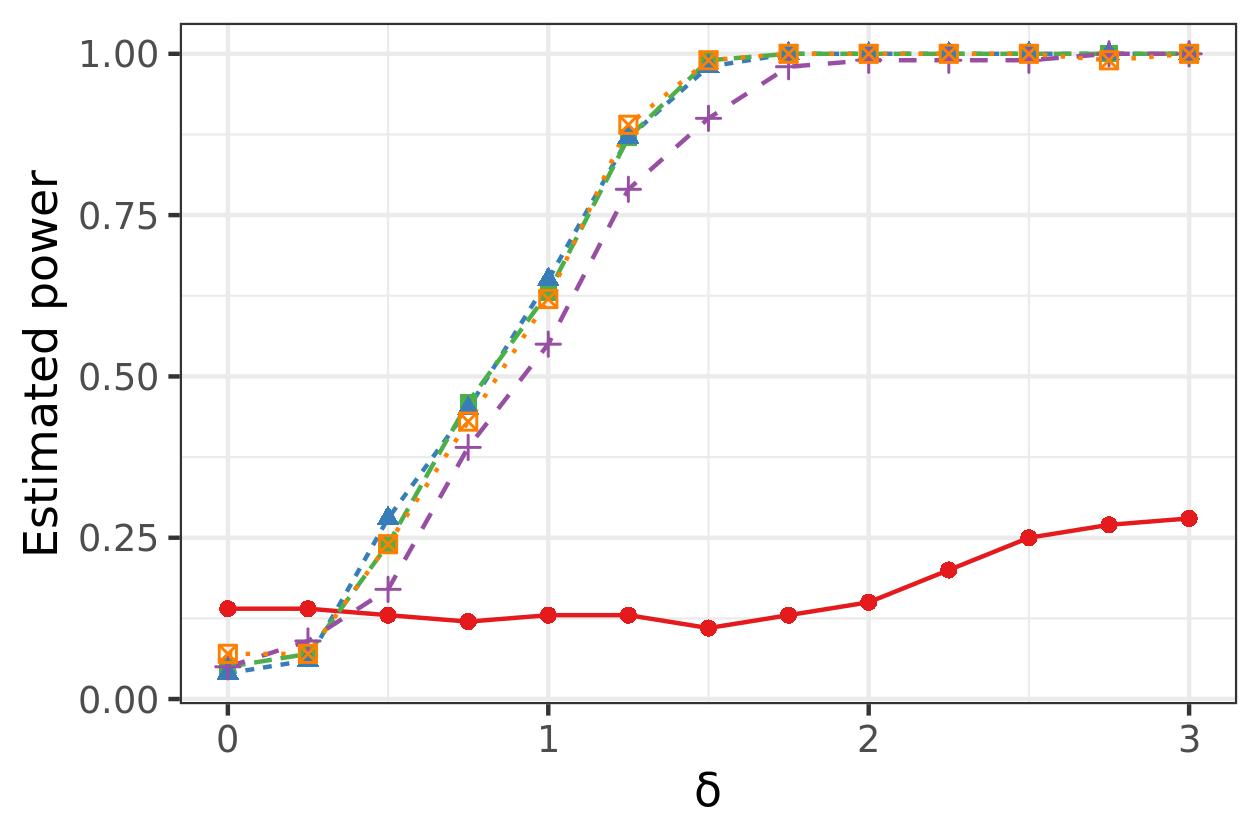}
	\end{minipage}
        }
        \subfigure{
          \rotatebox{90}{\scriptsize{~~~~~~~~~~~~~~Case 2}}
	\begin{minipage}[t]{0.3\textwidth}
		\includegraphics[width=\textwidth]{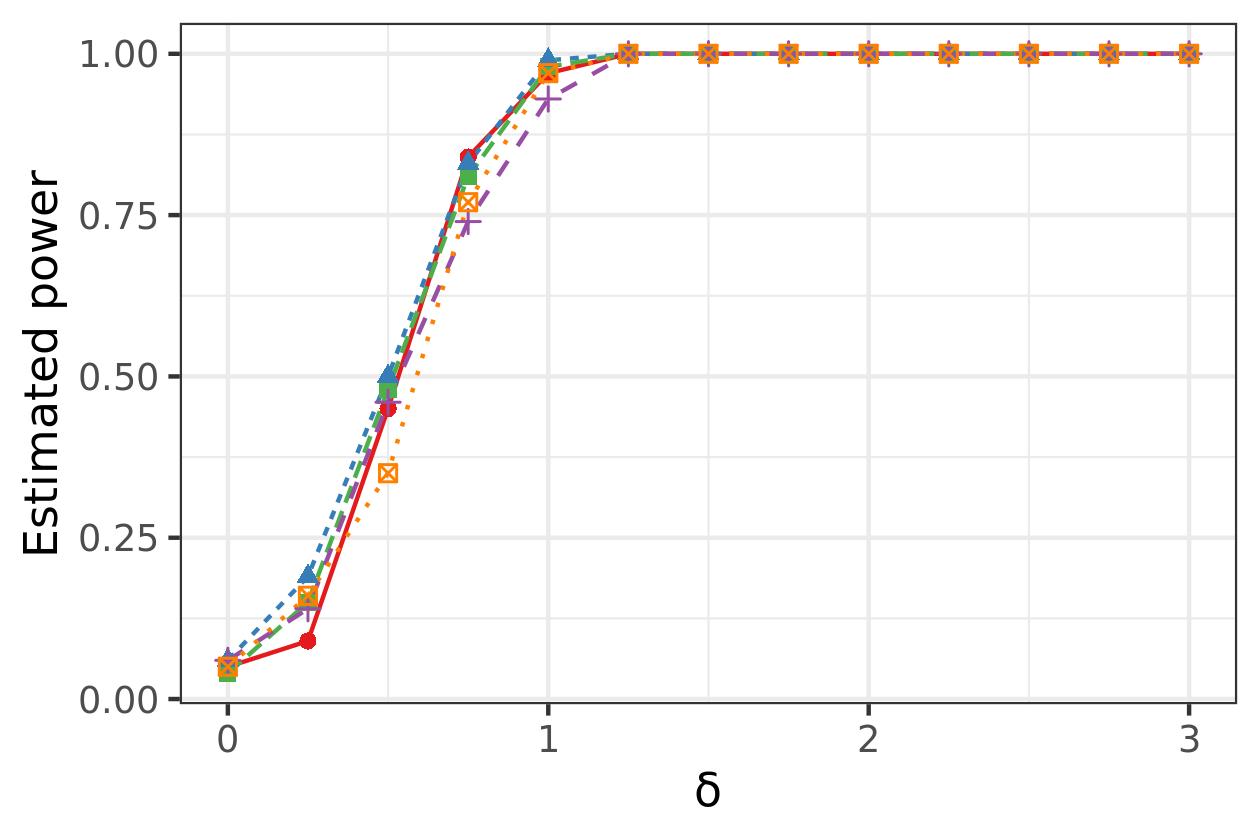}
	\end{minipage}
	}
	\subfigure{
        \begin{minipage}[t]{0.3\textwidth}
		\includegraphics[width=\textwidth]{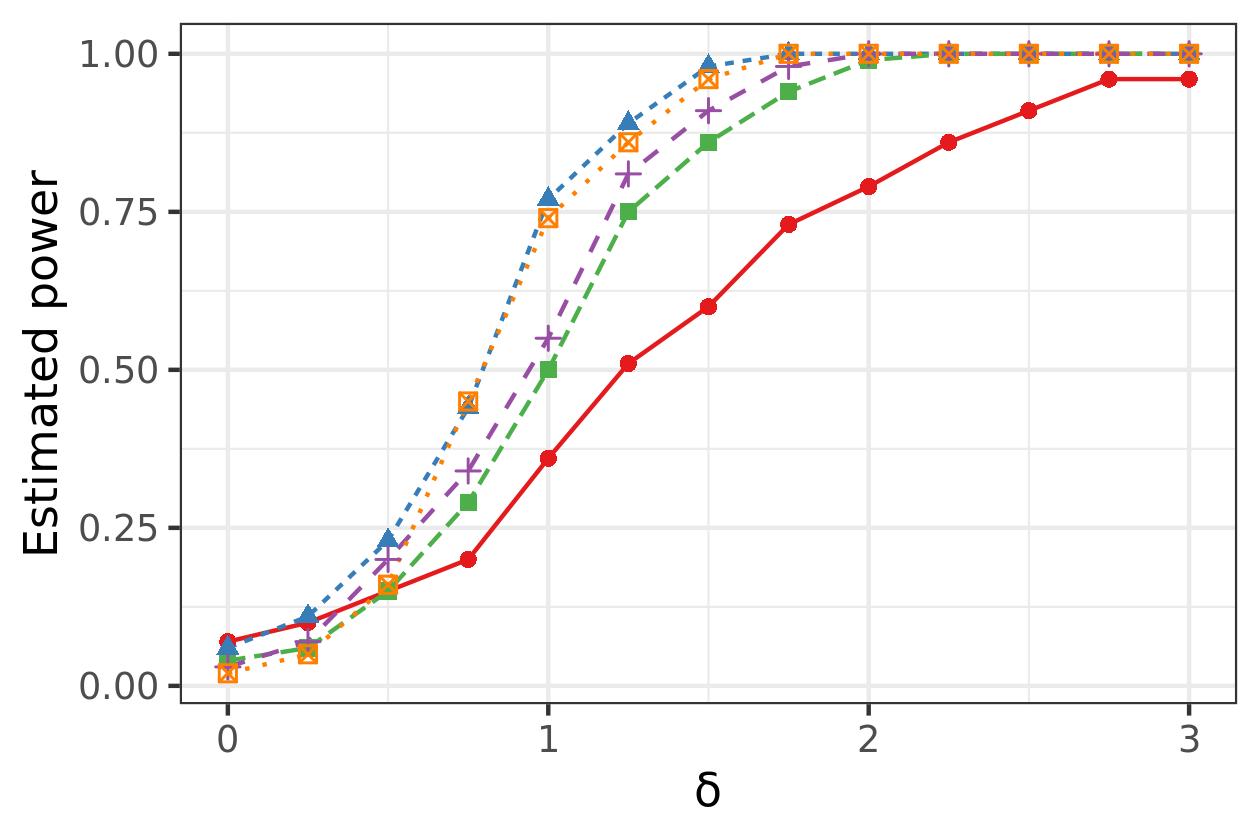}
	\end{minipage}
        }
        \subfigure{
        \begin{minipage}[t]{0.3\textwidth}
		\includegraphics[width=\textwidth]{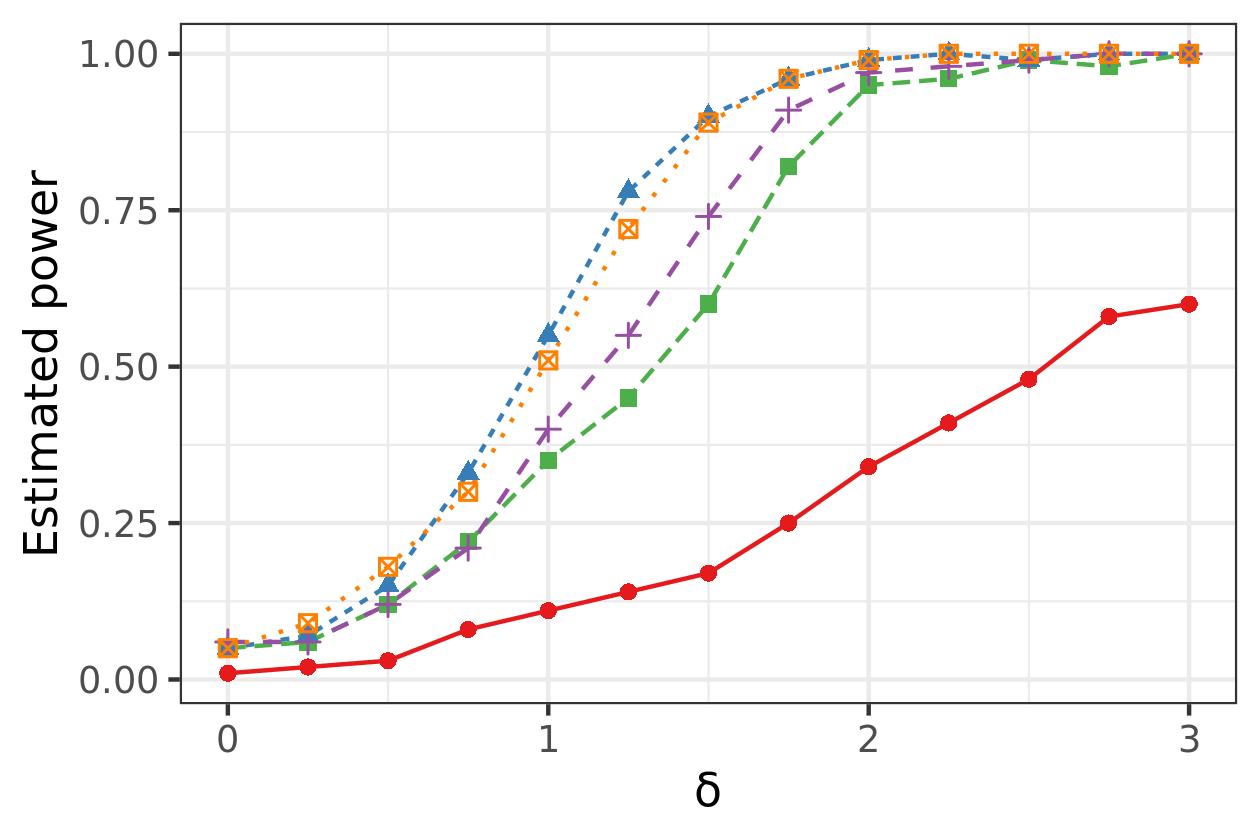}
	\end{minipage}
        }
        \subfigure{
          \rotatebox{90}{\scriptsize{~~~~~~~~~~~~~~Case 3}}
	\begin{minipage}[t]{0.3\textwidth}
		\includegraphics[width=\textwidth]{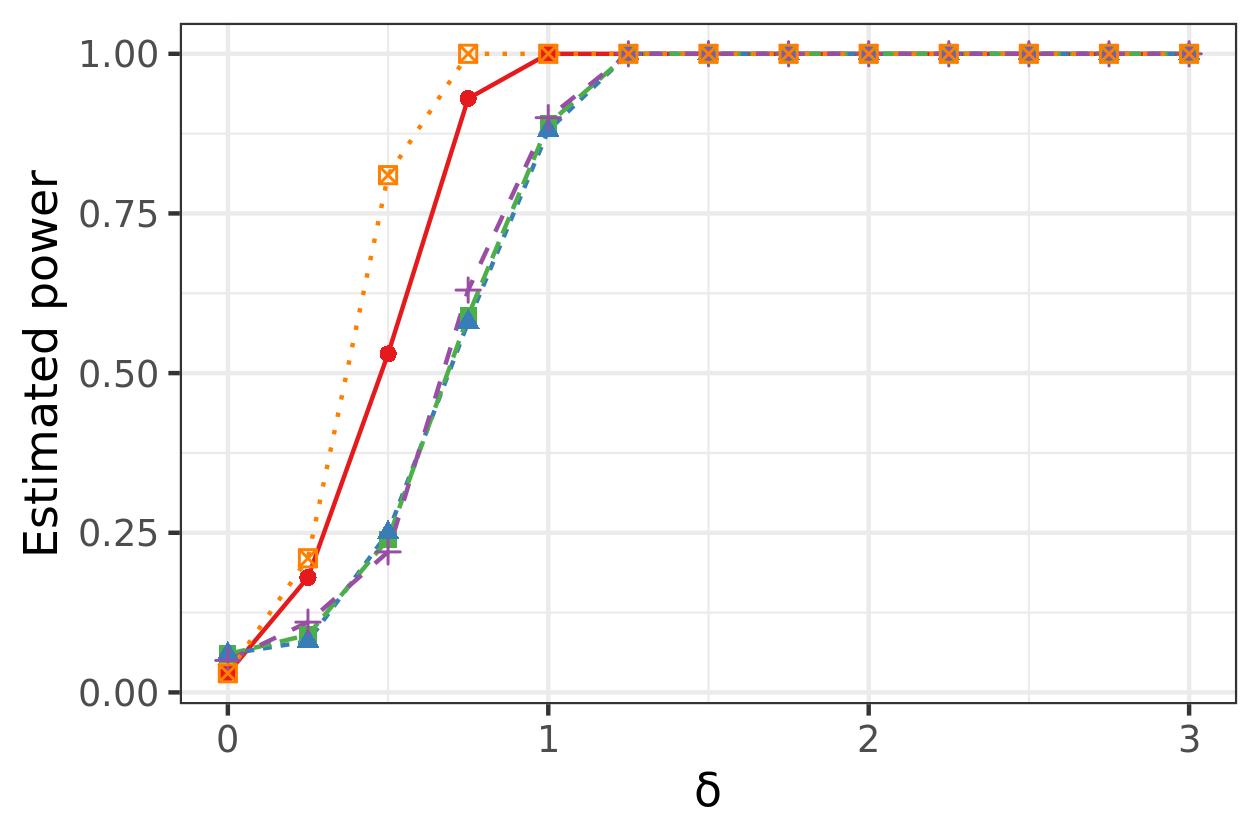}
	\end{minipage}
	}
	\subfigure{
        \begin{minipage}[t]{0.3\textwidth}
		\includegraphics[width=\textwidth]{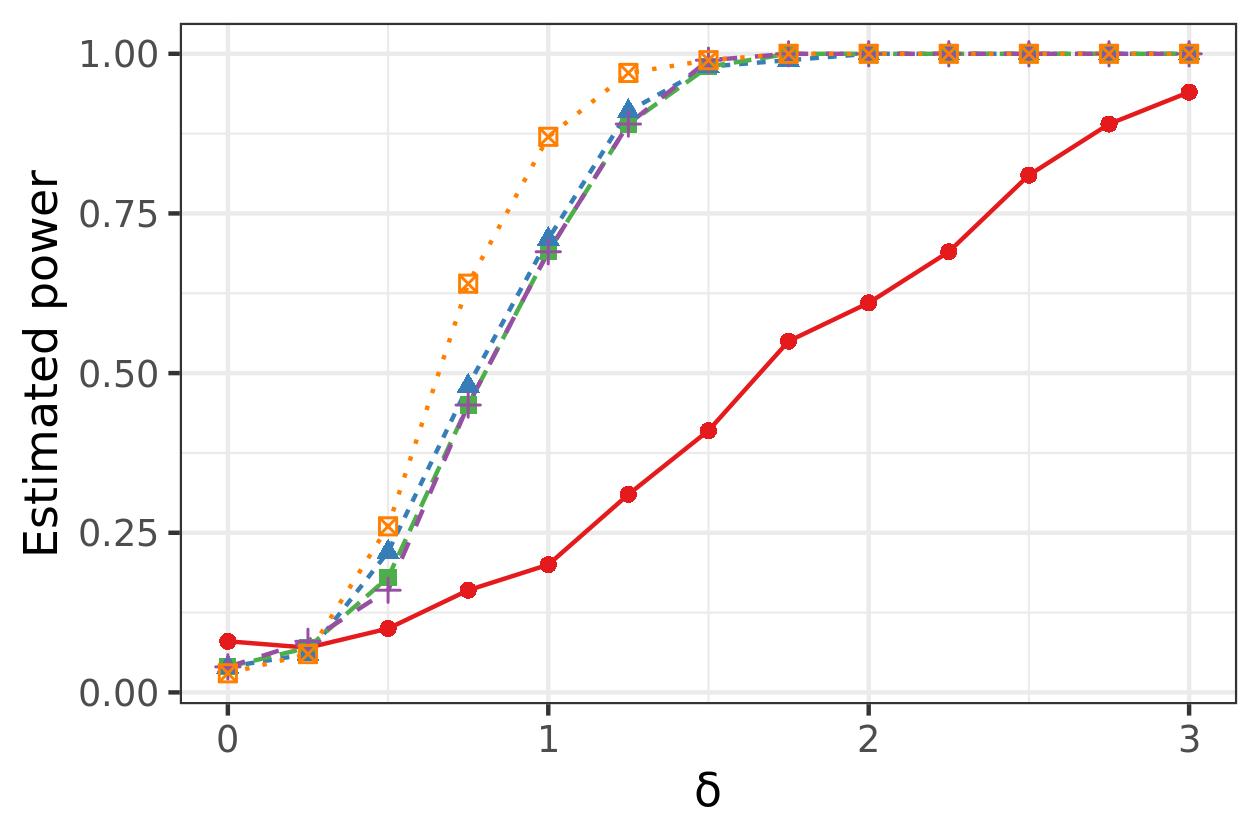}
	\end{minipage}
        }
        \subfigure{
        \begin{minipage}[t]{0.3\textwidth}
		\includegraphics[width=\textwidth]{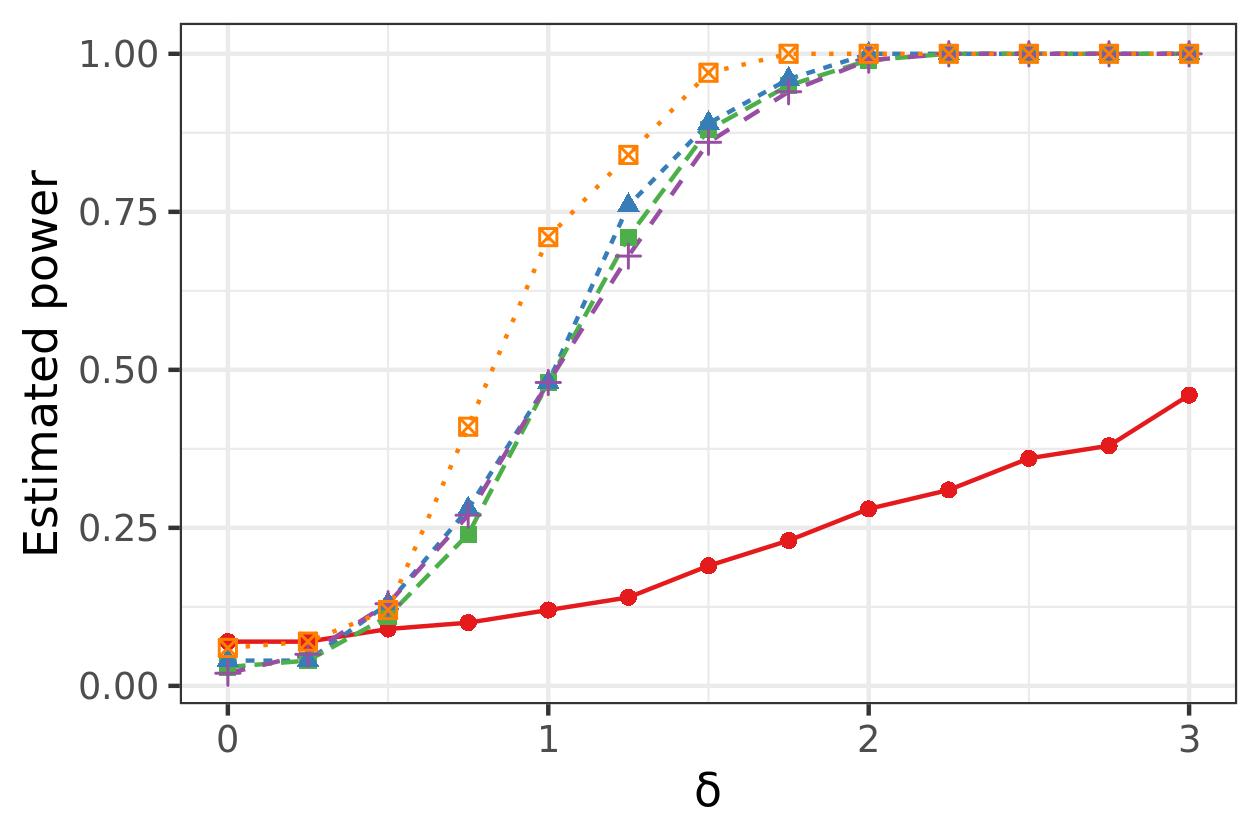}
	\end{minipage}
	}
        \setcounter{subfigure}{0}
        \subfigure[Scenario 1]{
          \rotatebox{90}{\scriptsize{~~~~~~~~~~~~~~Case 4}}
	\begin{minipage}[t]{0.3\textwidth}
		\includegraphics[width=\textwidth]{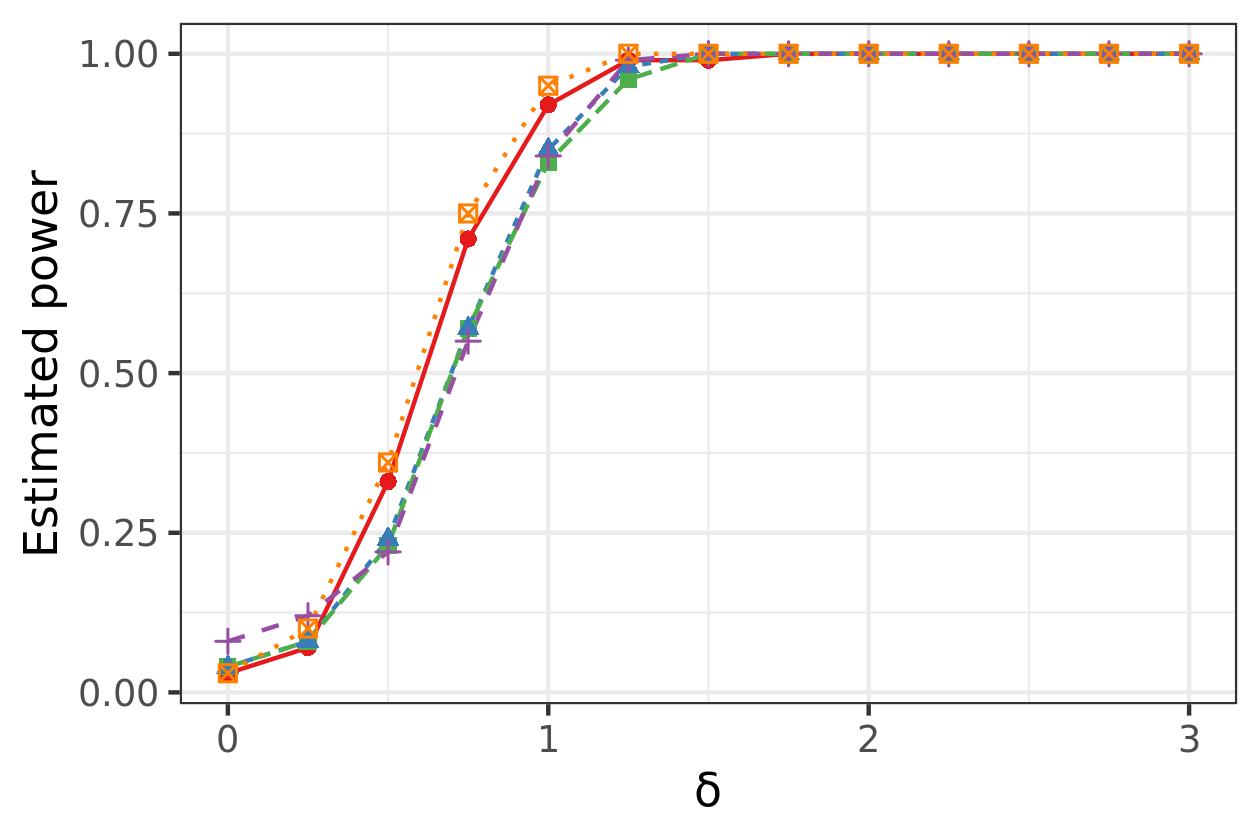}
	\end{minipage}
	}
	\subfigure[Scenario 2]{
	\begin{minipage}[t]{0.3\textwidth}
		\includegraphics[width=\textwidth]{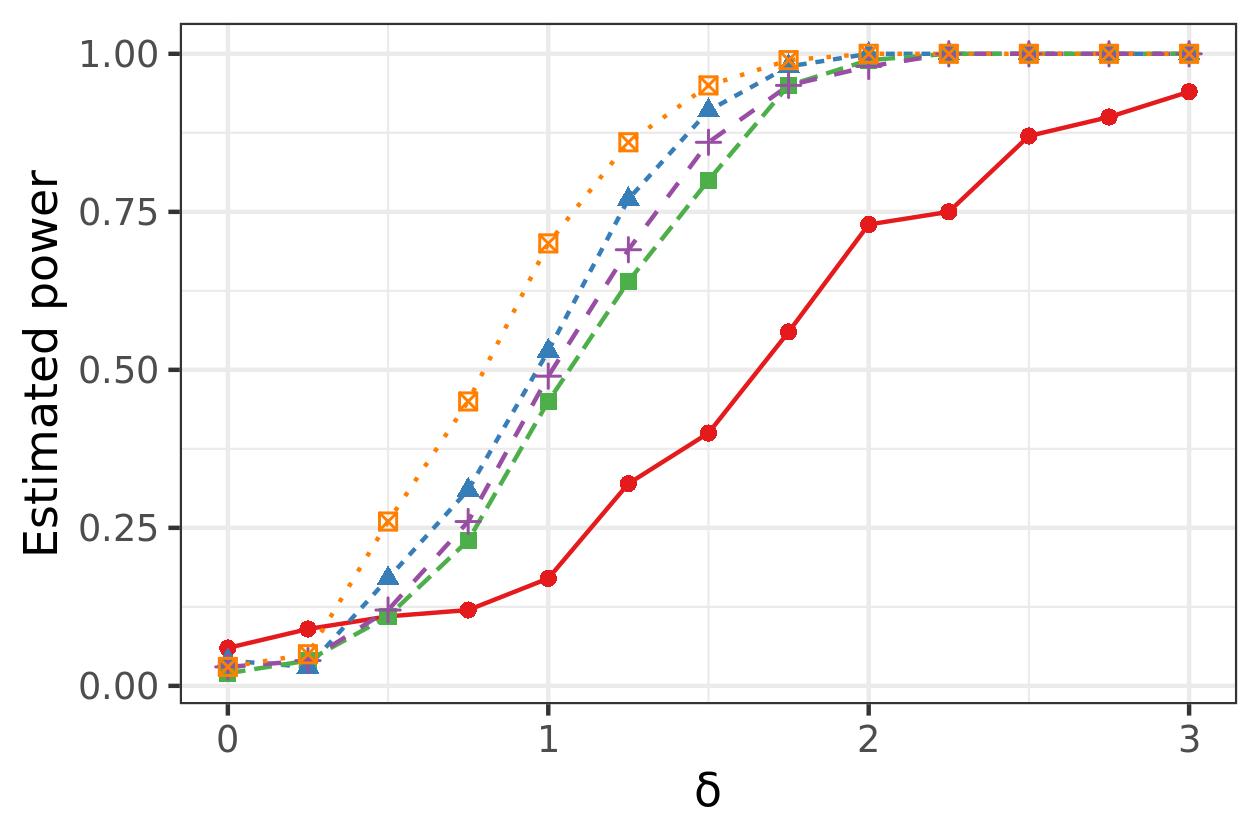}
	\end{minipage}
	}
        \subfigure[Scenario 3]{
	\begin{minipage}[t]{0.3\textwidth}
		\includegraphics[width=\textwidth]{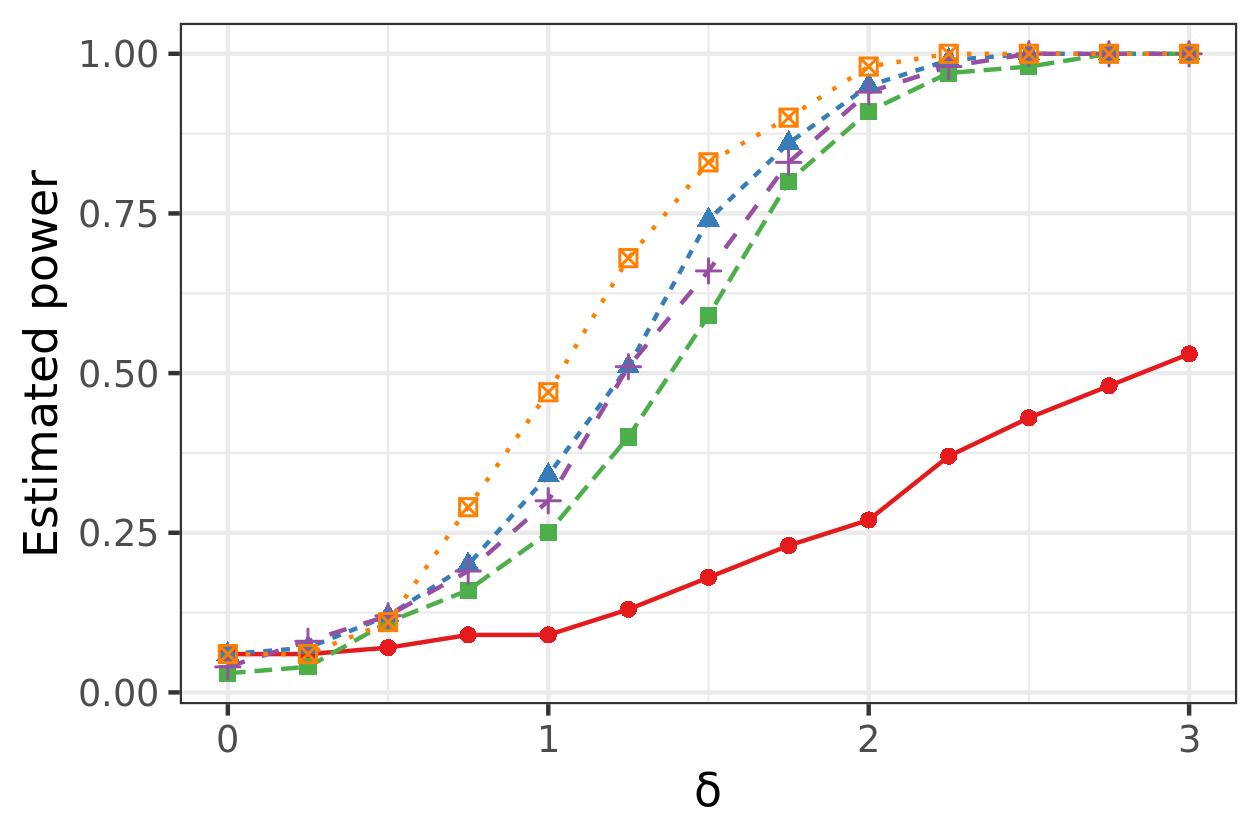}
	\end{minipage}
	}
	\caption{Estimated power curves under four different correlation structures of disturbance terms $\varepsilon(t)$'s (Case 1 to Case 4, from top to bottom) and three different types of sampling for time (Scenario 1 to Scenario 3, from left to right).}
        \label{fig: power curves}
\end{figure}

In this section, we evaluated the performance of our proposed permutation test for $H_{k0}$ and compared it with the FoSR test.

We focused on the candidate mediator $M_{6}$ with $\alpha_{6} = 0$ and $\beta_{6}(t) = 0.5 \delta \cos(\pi t)\mathbf{I}_{\{t \geq 0.5\} }$. Various values of $\delta$ were considered to evaluate the power of detecting mediation effects in the marginal time-varying coefficient model (\ref{Y_X+M_varying}). Smaller $\delta$ values generally result in lower power, especially with sparse data settings. The power is estimated as:
$$
\text{power} = \frac{\sum_{g=1}^{G} I\big( P_{\beta_{6}^{*}(t)}^{(g)} < \varrho \big) }{G},
$$ 
where $P_{\beta_{6}^{*}(t)}^{(g)}$ represents the permuted p-value at the $g$-th simulation replication.

The power was computed at a significance level of 0.05. The corresponding results, shown as power curves, under different setups are presented in Figure \ref{fig: power curves}. A higher power at a lower value of $\delta$ indicates a more effective testing procedure. In dense scenarios (Scenario 1), the performance of the FoSR test and the proposed methods is similar, but our method shows a higher power in sparse and irregularly spaced measured scenarios. Our method also demonstrates relatively consistent performance across different user-specified correlation structures, performing well regardless of whether the specified structure aligns with the actual data generation mechanism. In Scenario 3, the proposed methods significantly outperform the FoSR test, indicating that the pointwise test becomes less effective when the data is sparse and irregularly spaced. On the other hand, the proposed method remains relatively stable in different cases.

\subsection{The performance of mediator screening}
\label{subsec: The Performance of Mediator Screening}
\begin{figure}[]
	\centering
	\subfigure{
          \rotatebox{90}{\scriptsize{~~~~~~~~~~~~~~~FoSR}}
	\begin{minipage}[t]{0.30\textwidth}
		\includegraphics[width=\textwidth]{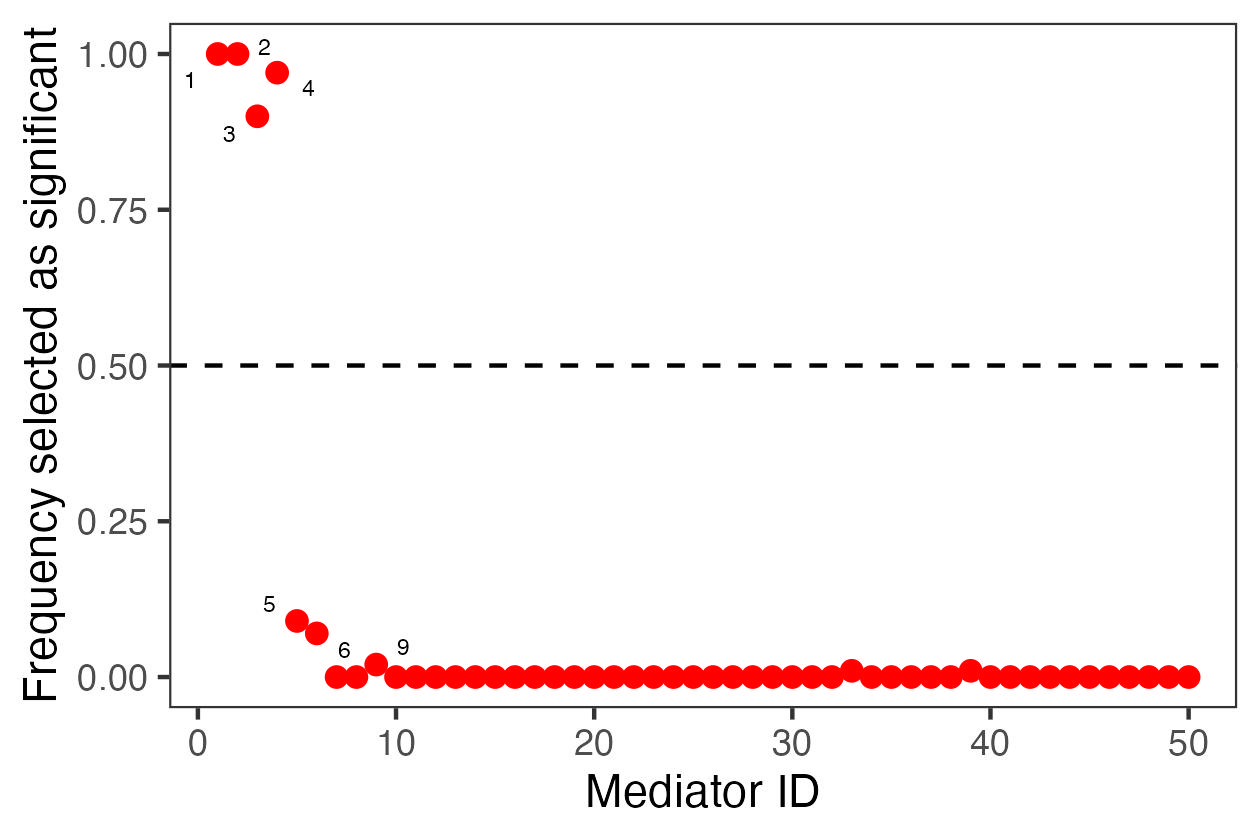}
	\end{minipage}
	}
	\subfigure{
        \begin{minipage}[t]{0.30\textwidth}
        \includegraphics[width=\textwidth]{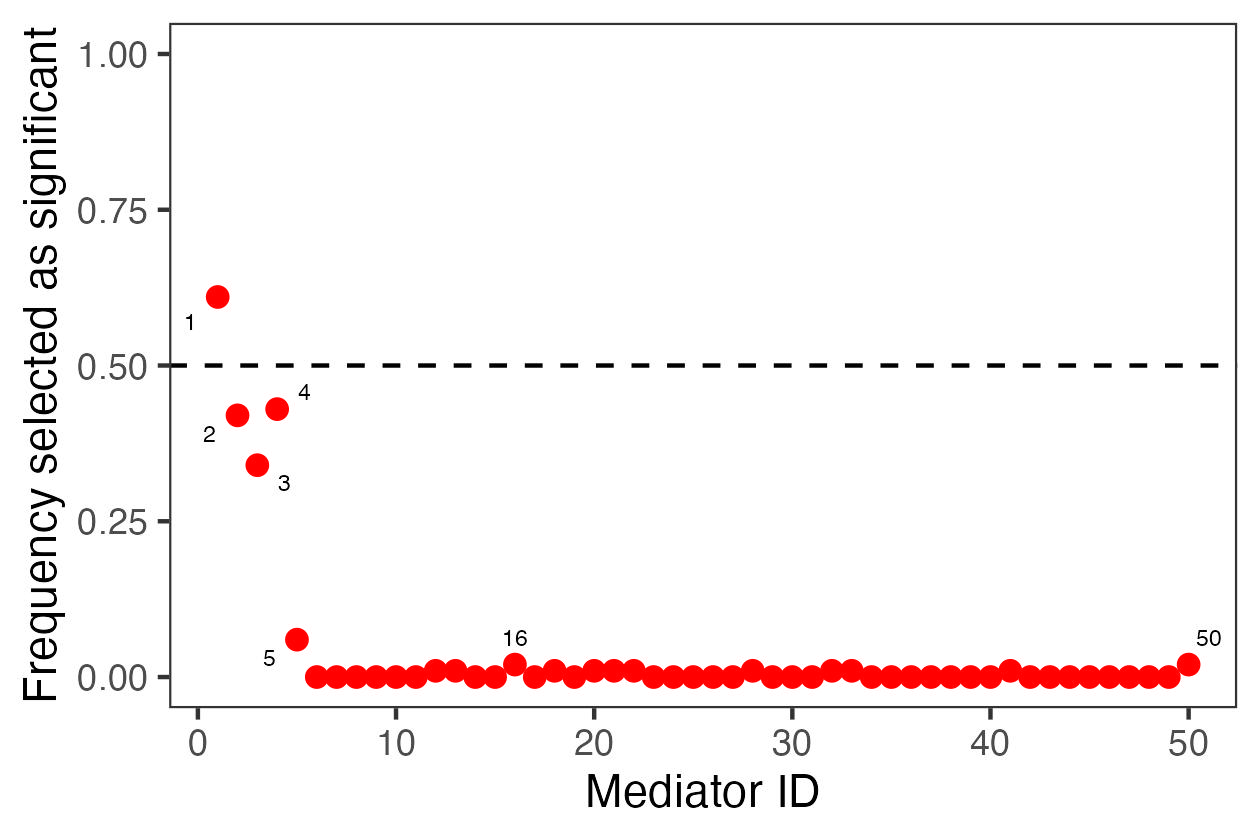}
	\end{minipage}
        }
        \subfigure{
        \begin{minipage}[t]{0.30\textwidth}
        \includegraphics[width=\textwidth]{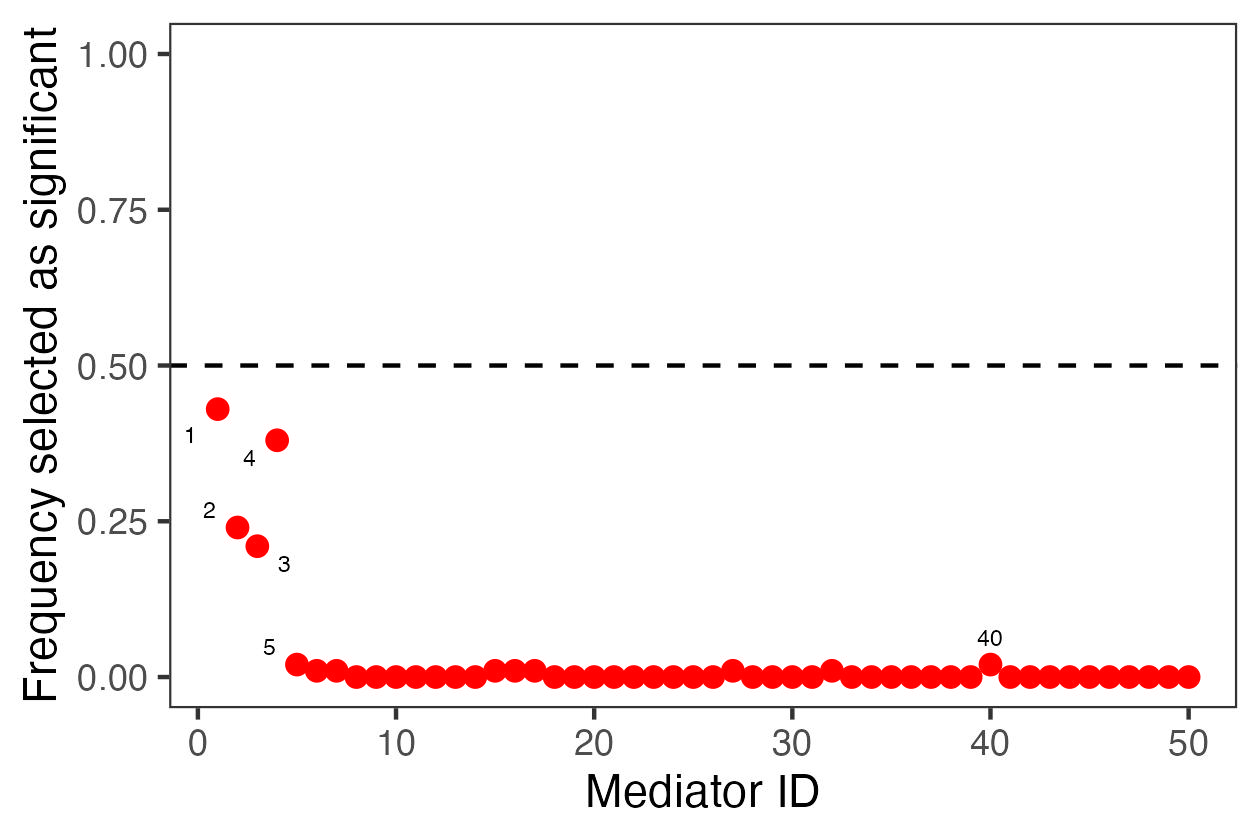}
	\end{minipage}
        }
        \subfigure{
          \rotatebox{90}{\scriptsize{~~~~~Proposed (Diagonal)}}
	\begin{minipage}[t]{0.30\textwidth}
		\includegraphics[width=\textwidth]{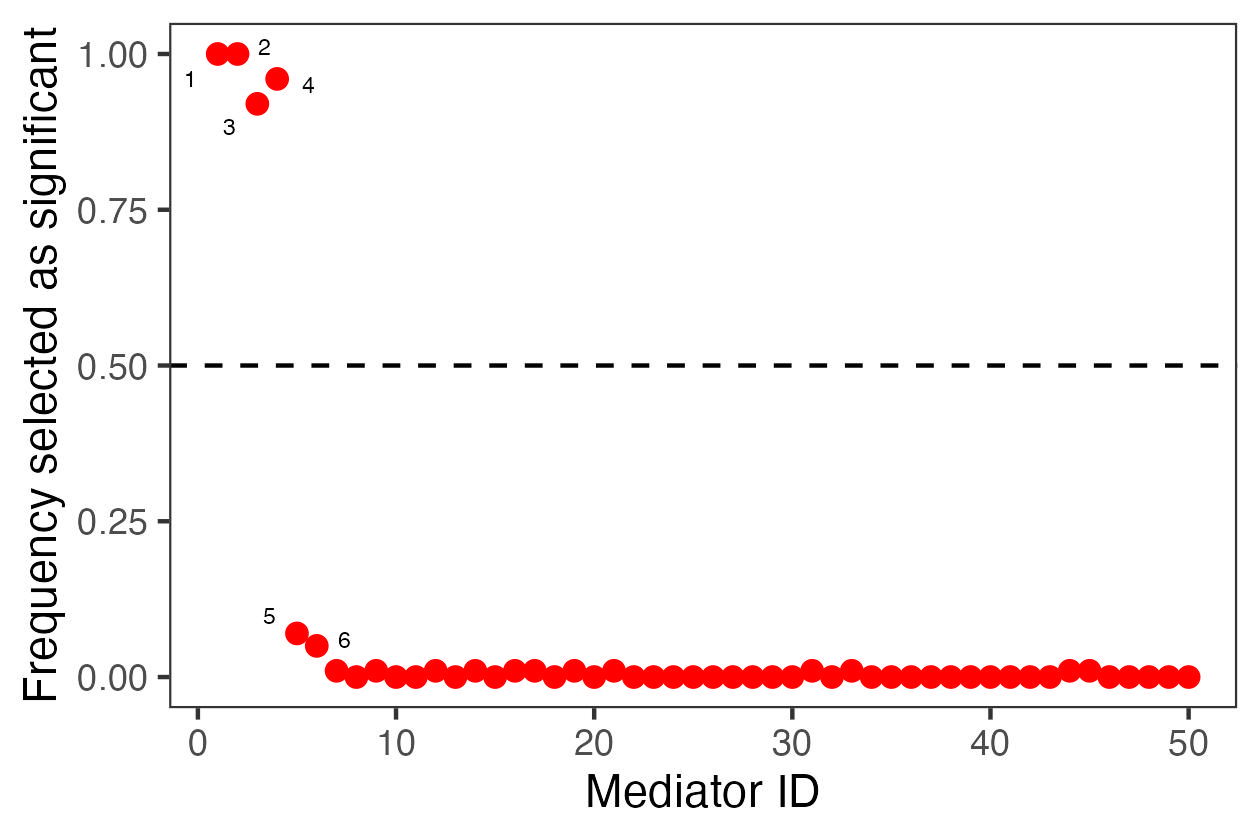}
	\end{minipage}
	}
	\subfigure{
        \begin{minipage}[t]{0.30\textwidth}
		\includegraphics[width=\textwidth]{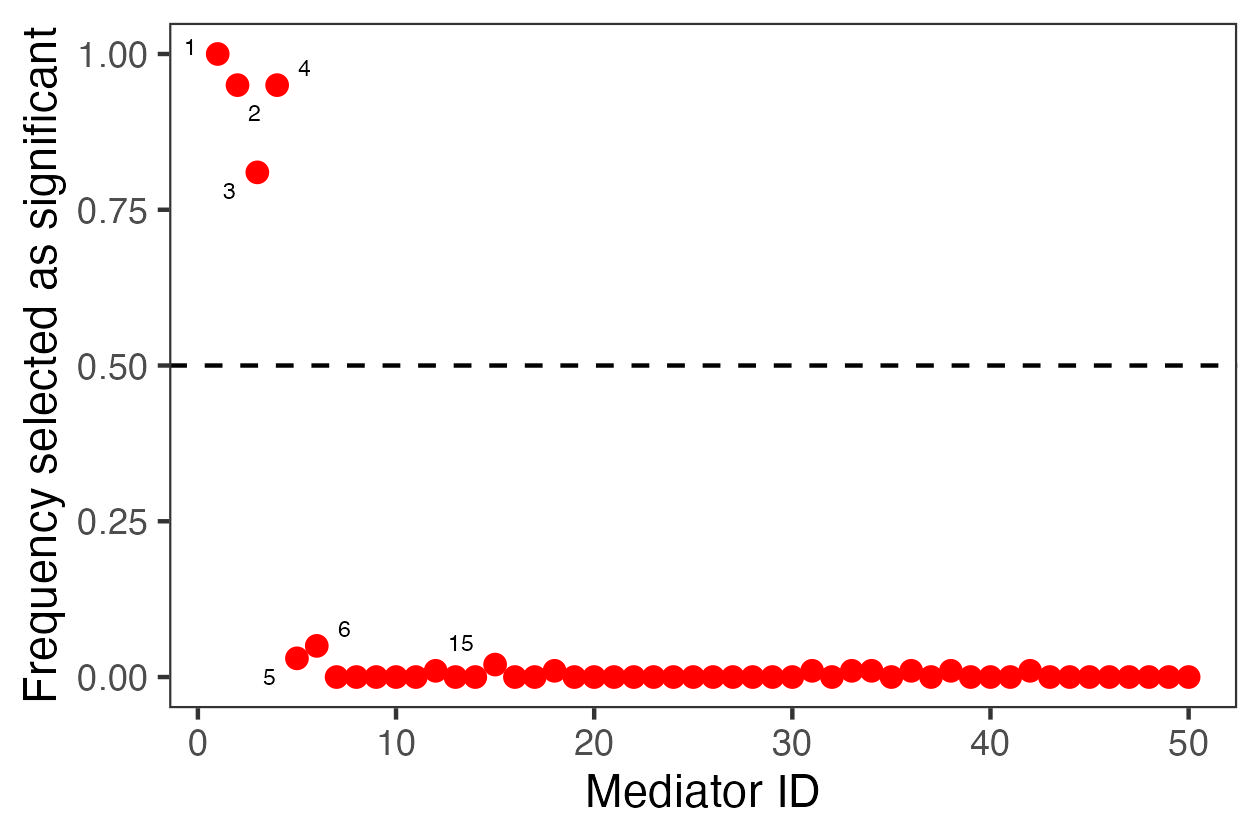}
	\end{minipage}
        }
        \subfigure{
        \begin{minipage}[t]{0.30\textwidth}
		\includegraphics[width=\textwidth]{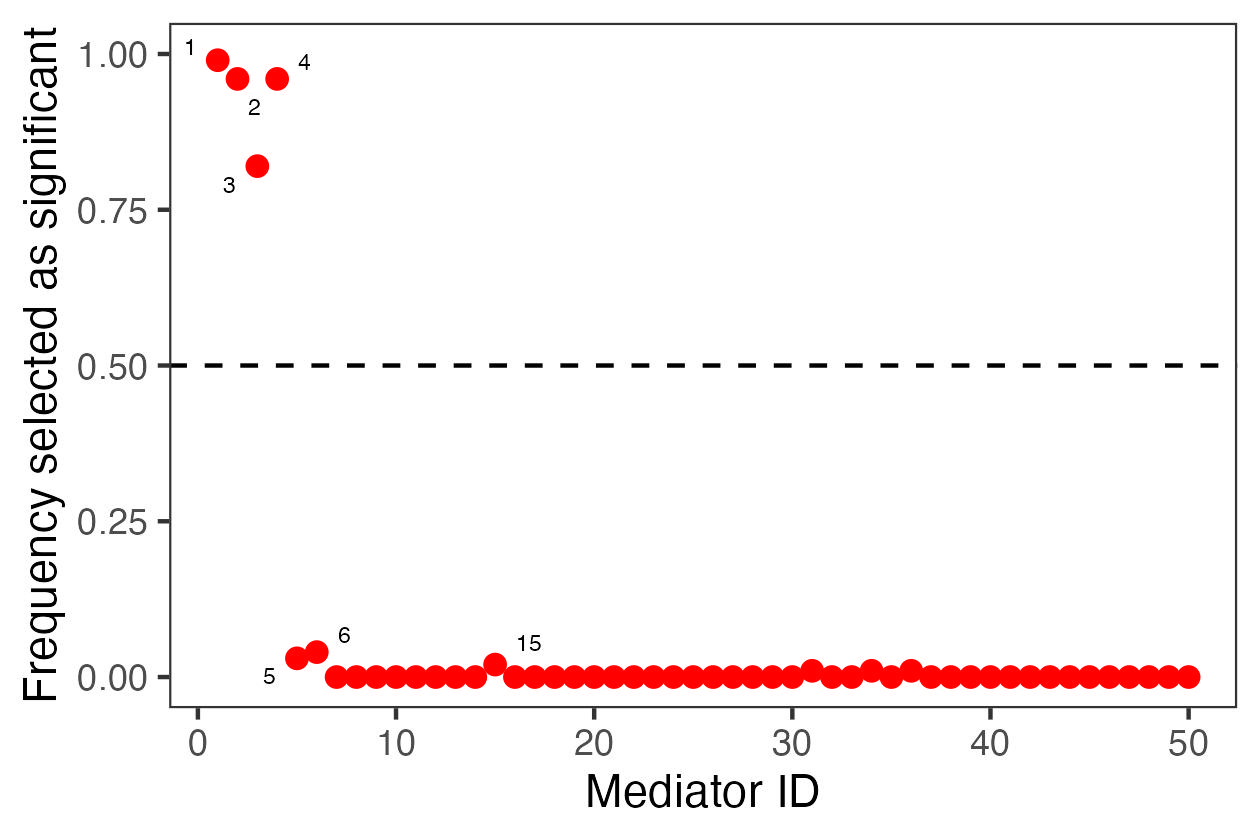}
	\end{minipage}
        }
        \subfigure{
          \rotatebox{90}{\scriptsize{~~~~~~~~~Proposed (AR)}}
	\begin{minipage}[t]{0.30\textwidth}
		\includegraphics[width=\textwidth]{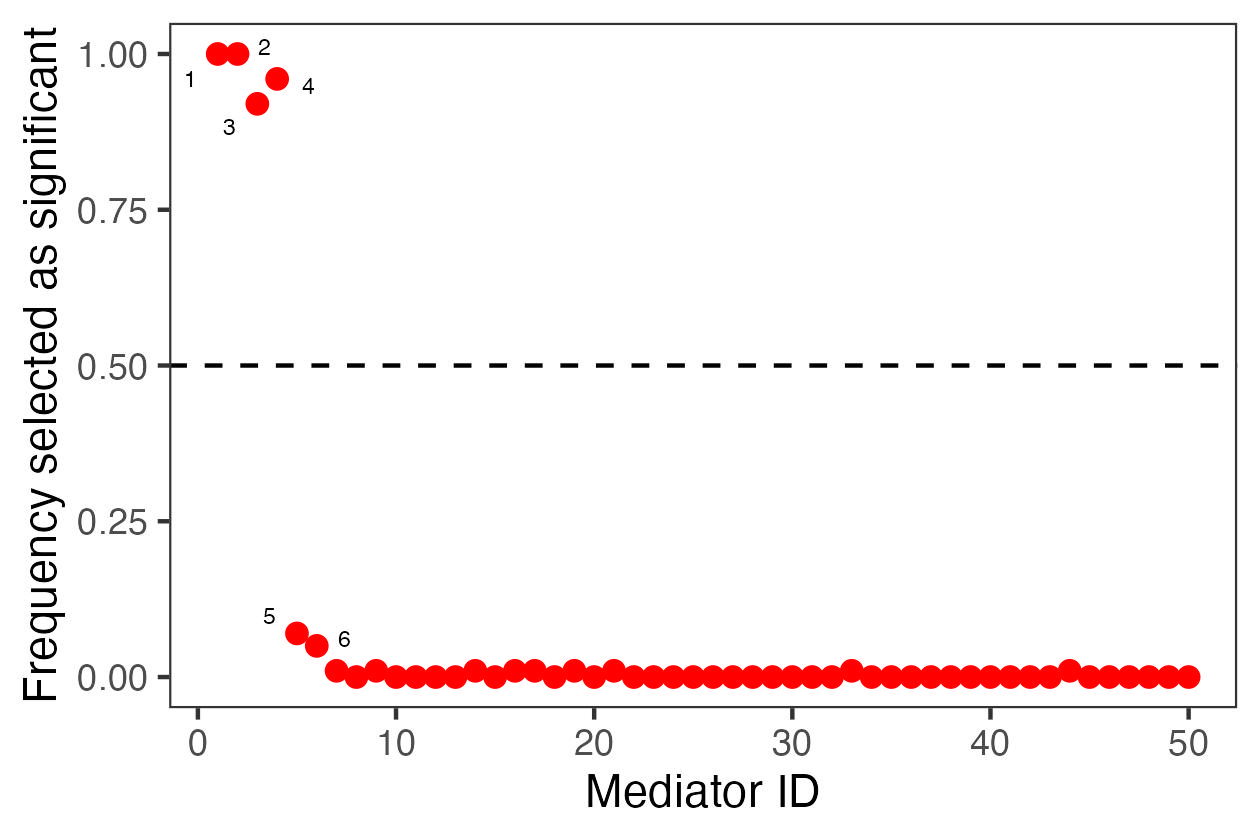}
	\end{minipage}
	}
	\subfigure{
        \begin{minipage}[t]{0.30\textwidth}
		\includegraphics[width=\textwidth]{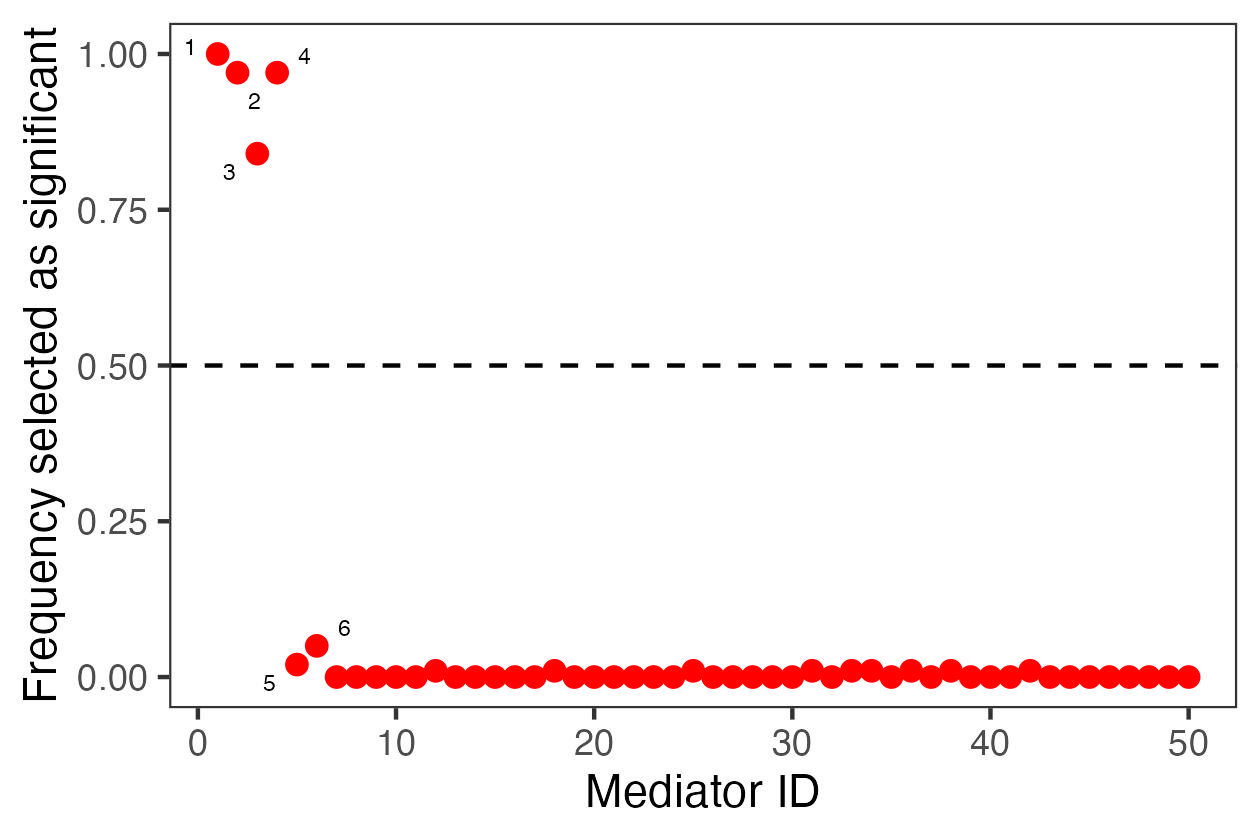}
	\end{minipage}
        }
        \subfigure{
        \begin{minipage}[t]{0.30\textwidth}
		\includegraphics[width=\textwidth]{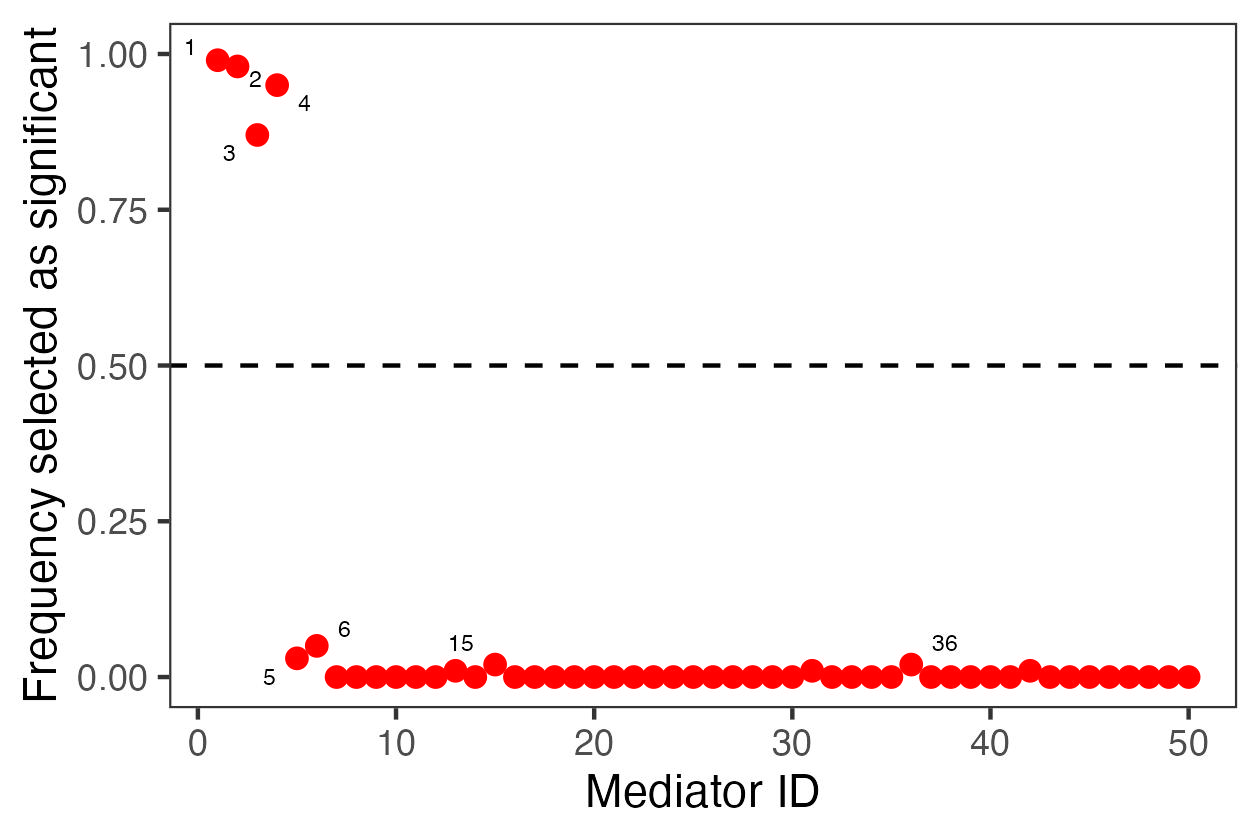}
	\end{minipage}
        }
        \subfigure{
          \rotatebox{90}{\scriptsize{~~~~~~~Proposed (Power)}}
	\begin{minipage}[t]{0.30\textwidth}
		\includegraphics[width=\textwidth]{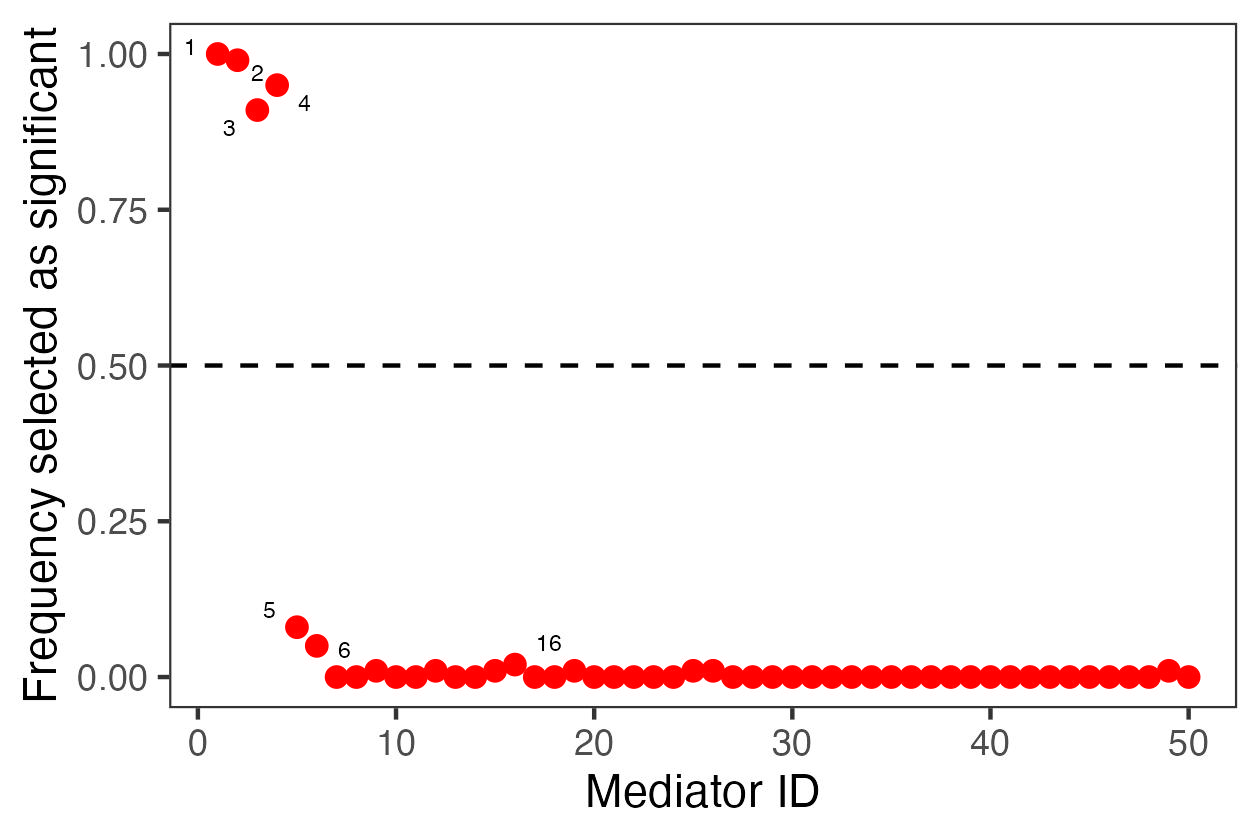}
	\end{minipage}
	}
	\subfigure{
        \begin{minipage}[t]{0.30\textwidth}
		\includegraphics[width=\textwidth]{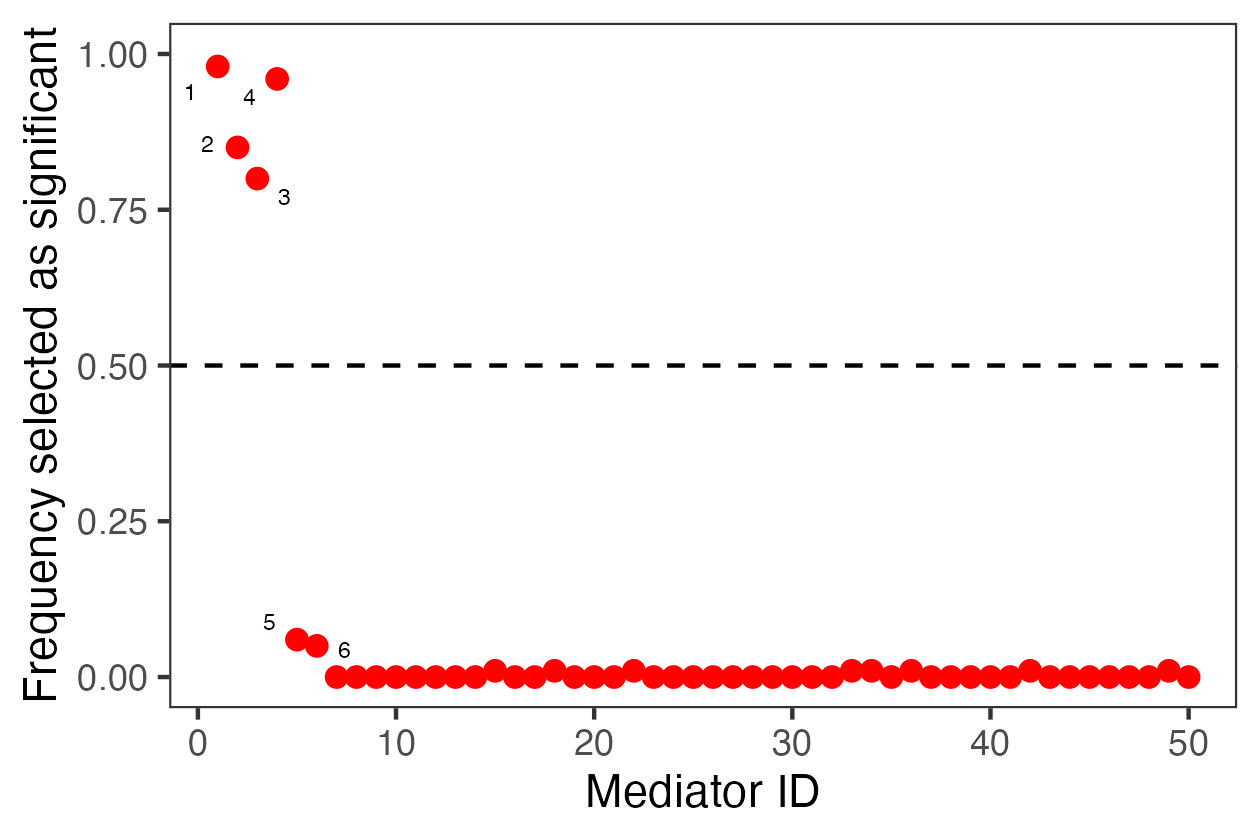}
	\end{minipage}
        }
        \subfigure{
        \begin{minipage}[t]{0.30\textwidth}
		\includegraphics[width=\textwidth]{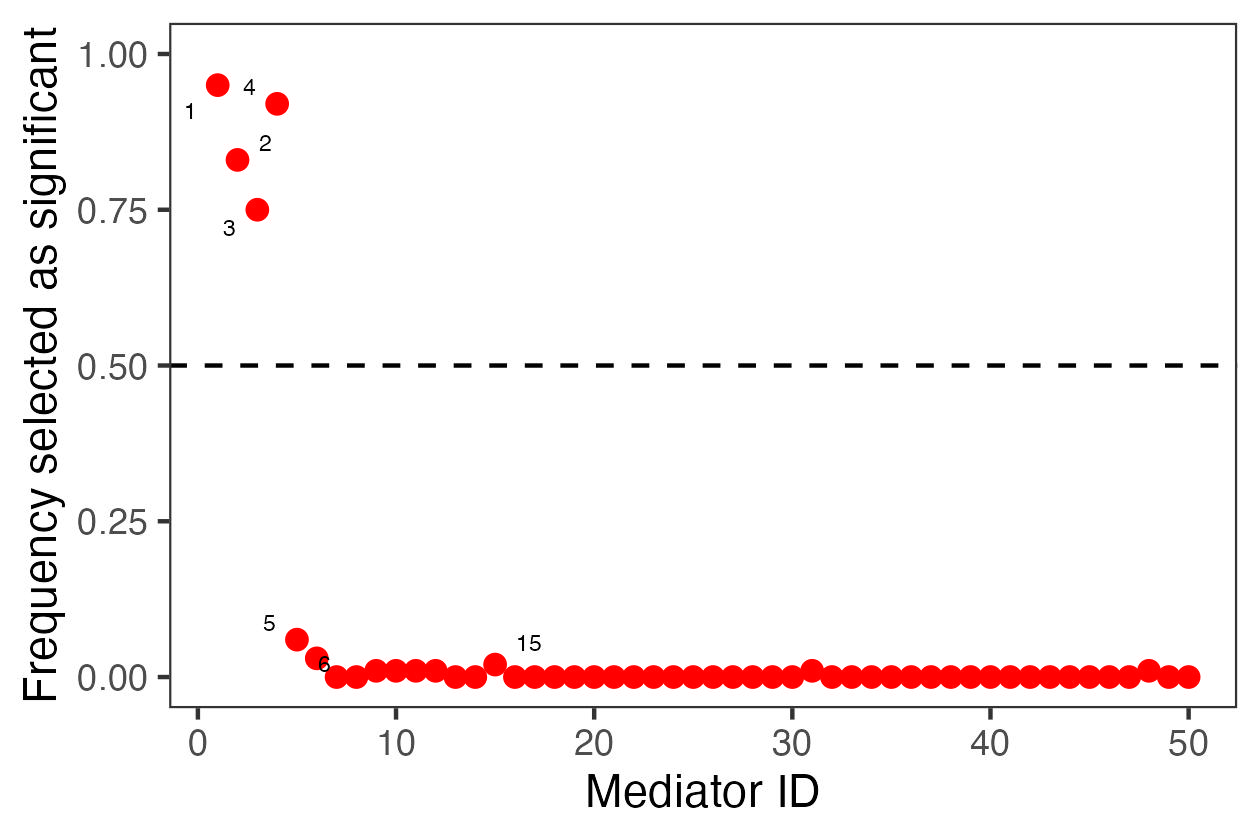}
	\end{minipage}
        }
        \setcounter{subfigure}{0}
        \subfigure[Scenario 1]{
          \rotatebox{90}{\scriptsize{~~~~~~Proposed (Uniform)}}
	\begin{minipage}[t]{0.30\textwidth}
		\includegraphics[width=\textwidth]{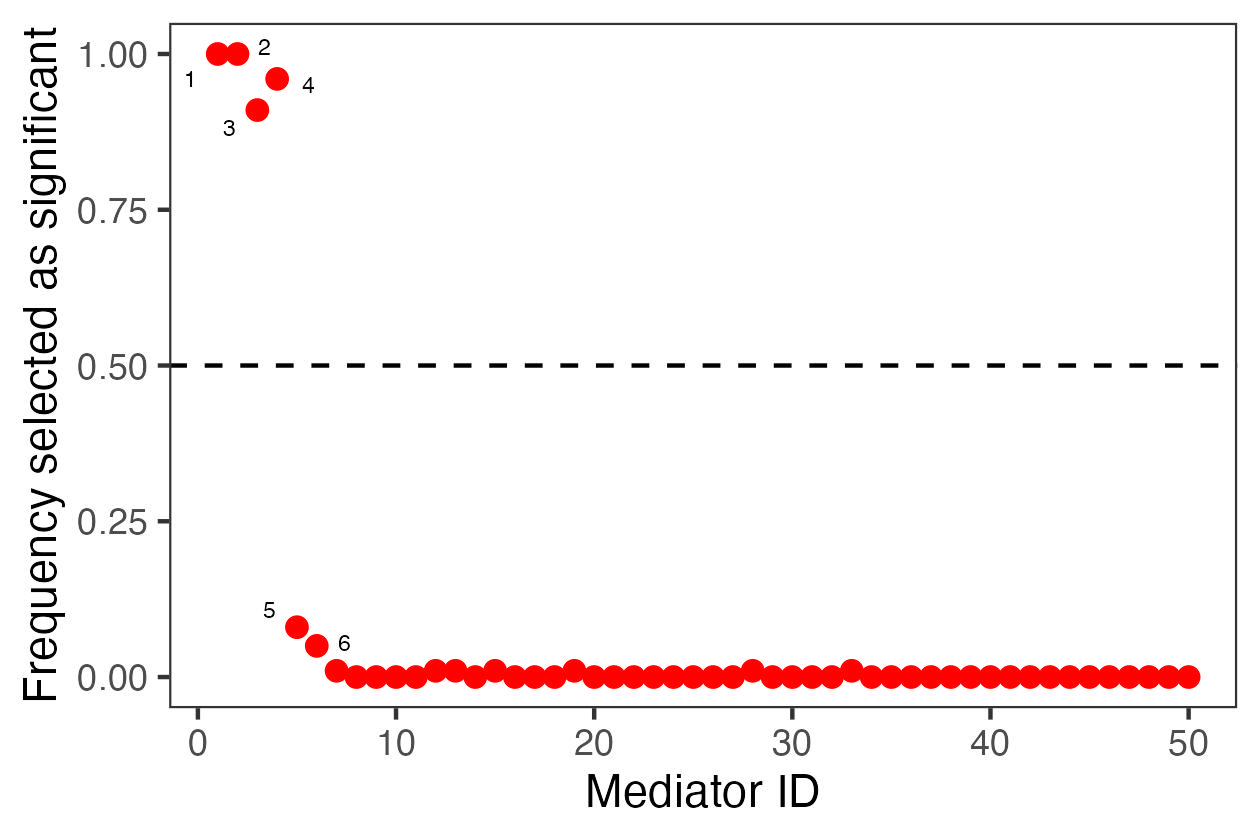}
	\end{minipage}
	}
	\subfigure[Scenario 2]{
	\begin{minipage}[t]{0.30\textwidth}
		\includegraphics[width=\textwidth]{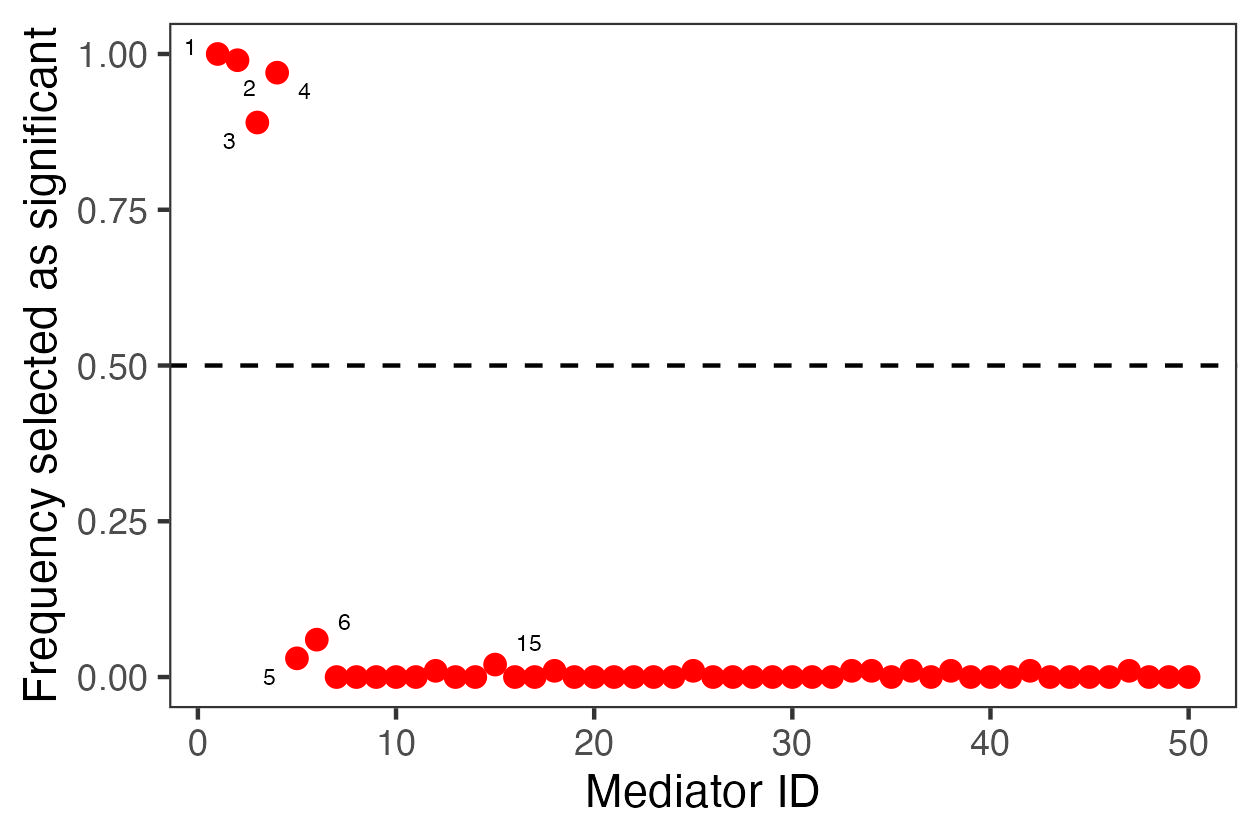}
	\end{minipage}
	}
        \subfigure[Scenario 3]{
	\begin{minipage}[t]{0.30\textwidth}
		\includegraphics[width=\textwidth]{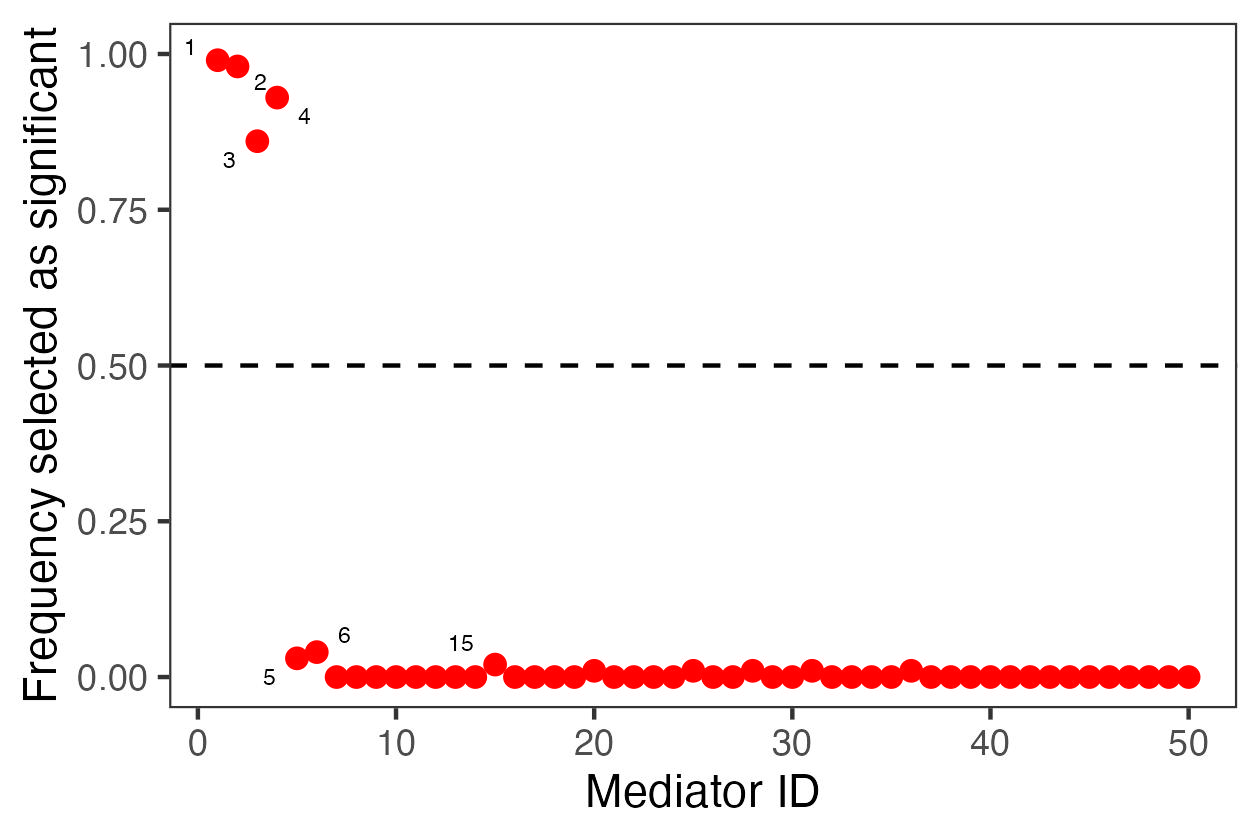}
	\end{minipage}
	}
	\caption{Results of estimated frequencies for different mediators (No.1 to No.50) under Case 1, using five different methods and three types of sampling for time (presented in rows and columns). }
        \label{fig: frequencies under Case 1}
\end{figure}

In this section, we evaluated the performance of our proposed methods for identifying significant mediators and compared them with the FoSR method. 

We continued with the data settings described in Section \ref{subsec: Design of Study} and focused on assessing whether the true mediators (mediators 1 through 4), for which $\alpha_{k}\beta_{k}(t) \neq 0$, could be successfully identified. The effectiveness of each mediator was evaluated by examining the frequency with which it was screened out across 100 simulation replications. A higher screening frequency reflects a stronger ability to detect a non-zero mediation effect for that mediator. The results for Case 1 are summarized in Figure \ref{fig: frequencies under Case 1}, and additional results for other cases are provided in Section S4 of the supplementary file.

For both the FoSR method and all variations of our proposed method, the screening frequencies for mediators 1 to 4 were close to 1 when the data were sufficiently dense (Scenario 1), demonstrating strong identification performance. However, under sparse measurement settings (Scenarios 2 and 3), the comparison method (FoSR) struggled to effectively detect the true mediators, with screening frequencies around or below 0.5, suggesting random identification. In contrast, our proposed methods exhibited relatively stable and higher screening frequencies in Scenarios 2 and 3, indicating greater robustness to data sparsity.

\subsection{The performance of mediation effect estimation}
\label{subsec: The performance of mediation effect estimation}

\begin{table}[]
  \centering
  \setlength{\tabcolsep}{2.5pt}
  \caption{Summary of performance metrics for mediation effect estimation: bias and standard deviation (sd) across different cases and scenarios. Subscripts ``i" and ``d" denote indirect and direct effects, respectively.}
    \begin{tabular}{lcccccccccccccc}
    \hline
          & \multicolumn{4}{c}{Scenario 1} &       & \multicolumn{4}{c}{Scenario 2} &       & \multicolumn{4}{c}{Scenario 3} \\
\cline{2-5}\cline{7-10}\cline{12-15}          & $\text{bias}_{\text{i}}$ & $\text{sd}_{\text{i}}$ & $\text{bias}_{\text{d}}$ & $\text{sd}_{\text{d}}$ &       & $\text{bias}_{\text{i}}$ & $\text{sd}_{\text{i}}$ & $\text{bias}_{\text{d}}$ & $\text{sd}_{\text{d}}$ &       & $\text{bias}_{\text{i}}$ & $\text{sd}_{\text{i}}$ & $\text{bias}_{\text{d}}$ & $\text{sd}_{\text{d}}$ \\
    \hline
    \textbf{Case 1} &       &       &       &       &       &       &       &       &       &       &       &       &       &  \\
    FoSR  & 0.06  & 0.45  & 0.03  & 0.64  &       & 0.06  & 0.45  & 0.03  & 0.64  &       & 0.06  & 0.45  & 0.03  & 0.64 \\
    Proposed (Diagonal) & 0.06  & 0.10  & 0.01  & 0.12  &       & 0.06  & 0.10  & 0.01  & 0.12  &       & 0.06  & 0.10  & 0.01  & 0.12 \\
    Proposed (AR) & 0.06  & 0.10  & 0.01  & 0.12  &       & 0.06  & 0.10  & 0.01  & 0.12  &       & 0.06  & 0.10  & 0.01  & 0.12 \\
    Proposed (Power) & 0.05  & 0.11  & 0.01  & 0.14  &       & 0.05  & 0.11  & 0.01  & 0.14  &       & 0.05  & 0.11  & 0.01  & 0.14 \\
    Proposed (Uniform) & 0.06  & 0.10  & 0.01  & 0.12  &       & 0.06  & 0.10  & 0.01  & 0.12  &       & 0.06  & 0.10  & 0.01  & 0.12 \\
    \textbf{Case 2} &       &       &       &       &       &       &       &       &       &       &       &       &       &  \\
    FoSR  & 0.05  & 0.28  & 0.02  & 0.41  &       & 0.05  & 0.28  & 0.02  & 0.41  &       & 0.05  & 0.28  & 0.02  & 0.41 \\
    Proposed (Diagonal) & 0.05  & 0.13  & 0.01  & 0.17  &       & 0.05  & 0.13  & 0.01  & 0.17  &       & 0.05  & 0.13  & 0.01  & 0.17 \\
    Proposed (AR) & 0.05  & 0.11  & 0.01  & 0.14  &       & 0.05  & 0.11  & 0.01  & 0.14  &       & 0.05  & 0.11  & 0.01  & 0.14 \\
    Proposed (Power) & 0.05  & 0.12  & 0.01  & 0.15  &       & 0.05  & 0.12  & 0.01  & 0.15  &       & 0.05  & 0.12  & 0.01  & 0.15 \\
    Proposed (Uniform) & 0.05  & 0.12  & 0.01  & 0.16  &       & 0.05  & 0.12  & 0.01  & 0.16  &       & 0.05  & 0.12  & 0.01  & 0.16 \\
    \textbf{Case 3} &       &       &       &       &       &       &       &       &       &       &       &       &       &  \\
    FoSR  & 0.06  & 0.35  & 0.03  & 0.57  &       & 0.06  & 0.35  & 0.03  & 0.57  &       & 0.06  & 0.35  & 0.03  & 0.57 \\
    Proposed (Diagonal) & 0.06  & 0.12  & 0.01  & 0.17  &       & 0.06  & 0.12  & 0.01  & 0.17  &       & 0.06  & 0.12  & 0.01  & 0.17 \\
    Proposed (AR) & 0.05  & 0.11  & 0.00  & 0.15  &       & 0.05  & 0.11  & 0.00  & 0.15  &       & 0.05  & 0.11  & 0.00  & 0.15 \\
    Proposed (Power) & 0.05  & 0.12  & 0.01  & 0.16  &       & 0.05  & 0.12  & 0.01  & 0.16  &       & 0.05  & 0.12  & 0.01  & 0.16 \\
    Proposed (Uniform) & 0.05  & 0.11  & 0.01  & 0.14  &       & 0.05  & 0.11  & 0.01  & 0.14  &       & 0.05  & 0.11  & 0.01  & 0.14 \\
    \textbf{Case 4} &       &       &       &       &       &       &       &       &       &       &       &       &       &  \\
    FoSR  & 0.05  & 0.33  & 0.03  & 0.51  &       & 0.05  & 0.33  & 0.03  & 0.51  &       & 0.05  & 0.33  & 0.03  & 0.51 \\
    Proposed (Diagonal) & 0.06  & 0.13  & 0.01  & 0.18  &       & 0.06  & 0.13  & 0.01  & 0.18  &       & 0.06  & 0.13  & 0.01  & 0.18 \\
    Proposed (AR) & 0.06  & 0.12  & 0.01  & 0.16  &       & 0.06  & 0.12  & 0.01  & 0.16  &       & 0.06  & 0.12  & 0.01  & 0.16 \\
    Proposed (Power) & 0.06  & 0.13  & 0.01  & 0.17  &       & 0.06  & 0.13  & 0.01  & 0.17  &       & 0.06  & 0.13  & 0.01  & 0.17 \\
    Proposed (Uniform) & 0.06  & 0.12  & 0.01  & 0.17  &       & 0.06  & 0.12  & 0.01  & 0.17  &       & 0.06  & 0.12  & 0.01  & 0.17 \\
    \hline
    \end{tabular}
  \label{tab: bias and sd}
\end{table}

To evaluate the estimation performance of the mediation effects, we assumed that the first $p_{0} = 4$ mediators were correctly identified in each simulation replicate. We utilized the natural direct effect and natural indirect effect estimators defined in equation (\ref{effect estimators}), and assessed estimation performance using bias and standard deviation (sd).

For any smooth function $h(t)$ over the domain $[0,T]$, the bias and sd are defined as:
\begin{align}
\text{bias} &=\left[ \int_{0}^{T} \left(E \widehat{h}^{(g)}\left(t \right)-h\left(t\right)\right)^{2} \mathrm{d}t \right]^{1 / 2}, \notag \\
\mathrm{sd} &=\left[\int_{0}^{T} E\left(\widehat{h}^{(g)}\left(t\right)-E \hat{h}^{(g)}\left(t\right)\right)^{2}\mathrm{d}t\right]^{1 / 2}, \notag
\end{align}
where $\widehat{h}^{(g)}(t)$ denotes the estimated function from the $g$-th simulation replicate, and $E\widehat{h}^{(g)}\left(t \right)$ is the average of $\widehat{h}^{(g)}(t)$ across all replicates, $g = 1, \ldots, 100$. For direct effects, we set $h(t) = \eta(t)$ and used its estimator in the $g$-th replicate, $\widehat{h}^{(g)}(t) = \boldsymbol{\psi}(t) \widehat{\boldsymbol{\zeta}}_{\eta}^{(g)}$, to compute the bias and standard deviation, denoted as $\text{bias}_{\text{d}}$ and $\text{sd}_{\text{d}}$, respectively. For indirect effects, we used $h(t) = \sum_{k=1}^{p_{0}}\alpha_{k}\beta_{k}(t)$ and its estimator in the $g$-th replicate, $\widehat{h}^{(g)}(t) = \sum_{k=1}^{p_{0}} \widehat{\alpha}_{k} \boldsymbol{\psi}(t) \widehat{\boldsymbol{\zeta}}_{\beta_{k}}^{(g)}$, to calculate the corresponding bias and standard deviation, denoted as $\text{bias}_{\text{i}}$ and $\text{sd}_{\text{i}}$. 

The estimation results are presented in Table \ref{tab: bias and sd}. These results indicate that the proposed methods achieve smaller bias and standard deviation compared to the alternative method, particularly in settings with sparse and irregularly spaced measurements. This demonstrates the competitive, and in some cases superior, performance of our proposed methods in estimating the mediation effect functions for the identified mediator variables.

\section{Case study}
\label{sec: Case Study}
In this section, we analyze the dataset described in Section \ref{sec: Motivation data} to identify lipid metabolic markers that link the acrophase RAR variable to cognitive decline through mediation pathways. 
\begin{figure}[]
    \centering
    \includegraphics[width=0.55\linewidth]{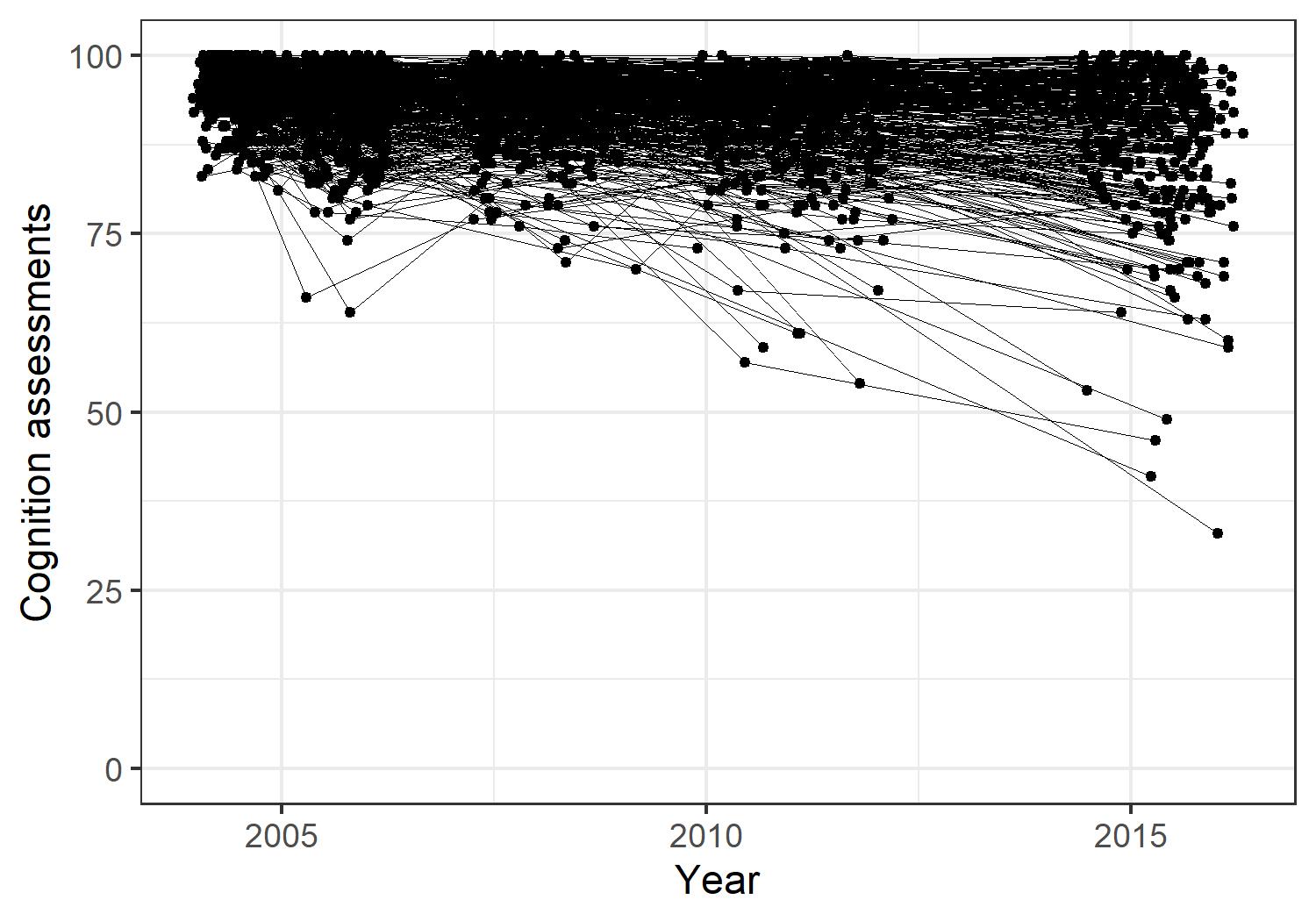}
    \caption{Individual-level cognitive scores over time.}
    \label{fig: Individual-level cognitive scores over time}
\end{figure}

The analysis included 490 participants and 476 lipid candidate mediators. Covariates ($\boldsymbol{Z}$) included baseline age (continuous), education level (continuous), hypertension (binary; 1 = yes, 0 = no), depression (binary; 1 = yes, 0 = no), cardiovascular disease (CVD; binary; 1 = yes, 0 = no), diabetes (binary; 1 = yes, 0 = no), self-rated health (binary; 1 = good or excellent, 0 = fair, poor, or very poor), body mass index (BMI, continuous), season (categorical, represented by three dummy variables for the four seasons), and study site (categorical, represented by five dummy variables corresponding to six locations: Birmingham, AL; Minneapolis, MN; Palo Alto, CA; Pittsburgh, PA; Portland, OR; San Diego, CA). As previously discussed, a challenge in analyzing the data is the relatively short follow-up period and irregular schedule. The trajectory is visualized in Figure \ref{fig: Individual-level cognitive scores over time}.

\begin{table}[]
  \centering
  \setlength{\tabcolsep}{2.75pt}
  \caption{Summary of results for selected mediators per methods. See Table 3 for detailed information of mediators that are consistently selected.}
    \begin{tabular}{clcccccccccccccc}
    \hline
          &       & \multicolumn{14}{c}{Number} \\
\cline{3-16}    FDR   & \multicolumn{1}{c}{Method} & 18    & 83    & 87    & 96    & 105   & 257   & 314   & 315   & 316   & 333   & 438   & 440   & 465   & 472 \\
    \hline
    $b=0.05$ & Proposed (Diagonal) &       & $\checkmark$ &       & $\checkmark$ &       &       &       &       &       &       & $\checkmark$ & $\checkmark$ &       & $\checkmark$ \\
          & Proposed (AR) &       & $\checkmark$ &       & $\checkmark$ &       &       &       &       &       &       & $\checkmark$ & $\checkmark$ &       & $\checkmark$ \\
          & Proposed (Power) &       & $\checkmark$ &       & $\checkmark$ &       &       &       &       &       &       & $\checkmark$ & $\checkmark$ &       & $\checkmark$ \\
          & Proposed (Uniform) &       & $\checkmark$ &       & $\checkmark$ &       &       &       &       &       &       & $\checkmark$ & $\checkmark$ &       & $\checkmark$ \\
          &       &       &       &       &       &       &       &       &       &       &       &       &       &       &  \\
    $b=0.10$ & Proposed (Diagonal) &       & $\checkmark$ & $\checkmark$ & $\checkmark$ & $\checkmark$ &       & $\checkmark$ & $\checkmark$ & $\checkmark$ & $\checkmark$ & $\checkmark$ & $\checkmark$ & $\checkmark$ & $\checkmark$ \\
          & Proposed (AR) &       & $\checkmark$ &       & $\checkmark$ &       &       &       &       &       &       & $\checkmark$ & $\checkmark$ &       & $\checkmark$ \\
          & Proposed (Power) &       & $\checkmark$ & $\checkmark$ & $\checkmark$ & $\checkmark$ &       & $\checkmark$ & $\checkmark$ & $\checkmark$ & $\checkmark$ & $\checkmark$ & $\checkmark$ & $\checkmark$ & $\checkmark$ \\
          & Proposed (Uniform) & $\checkmark$ & $\checkmark$ & $\checkmark$ & $\checkmark$ &       & $\checkmark$ &       &       & $\checkmark$ &       & $\checkmark$ & $\checkmark$ & $\checkmark$ & $\checkmark$ \\
    \hline
    \end{tabular}
    \label{tab: Summary results for selected mediators}
\end{table}

\begin{figure}[]
	\centering
	\subfigure[Direct effect.]{
        \begin{minipage}[t]{0.30\textwidth}
        \includegraphics[width=\textwidth]{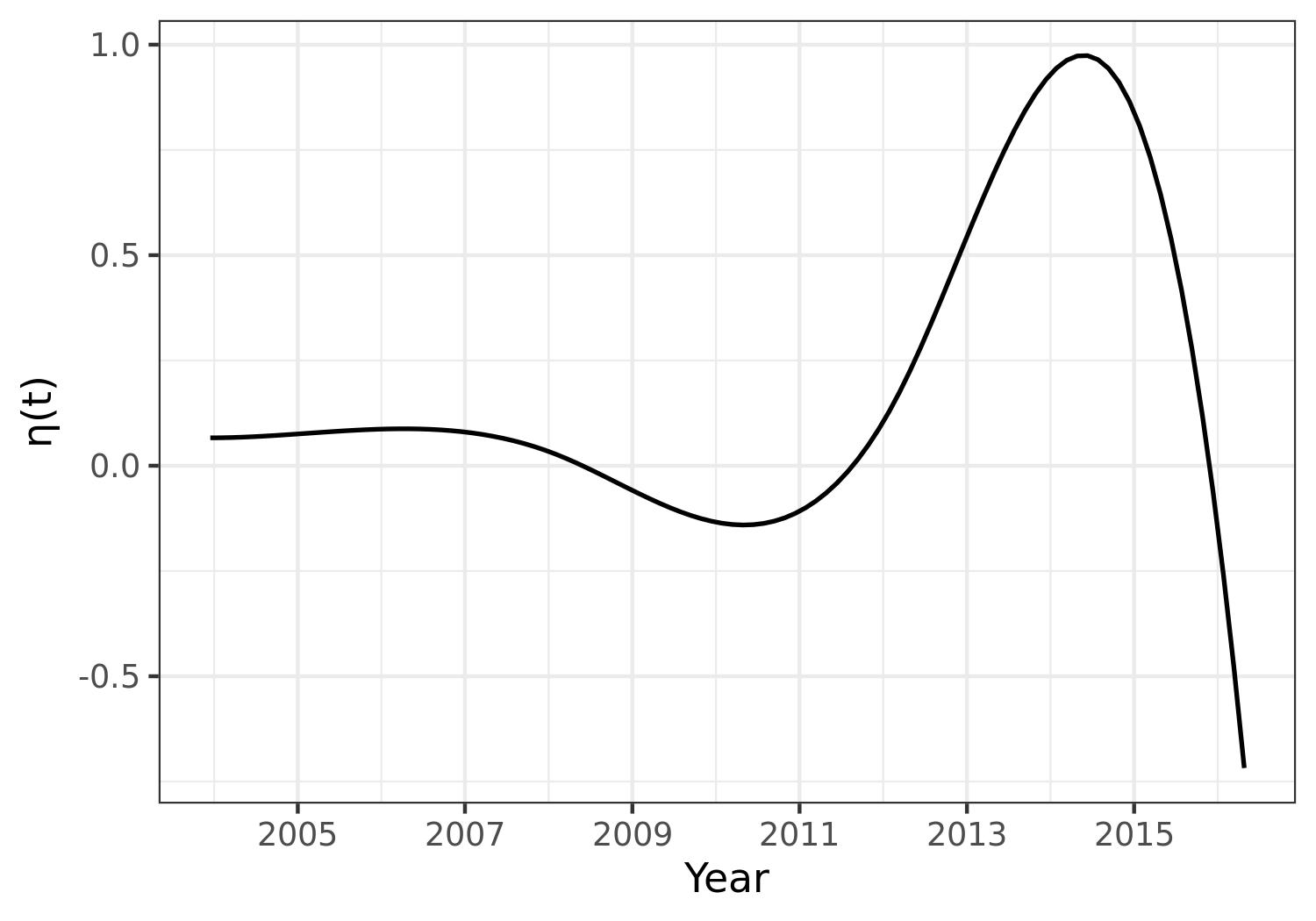}
	\end{minipage}
        }
        \subfigure[Indirect effects (No.83).]{
        \begin{minipage}[t]{0.30\textwidth}
        \includegraphics[width=\textwidth]{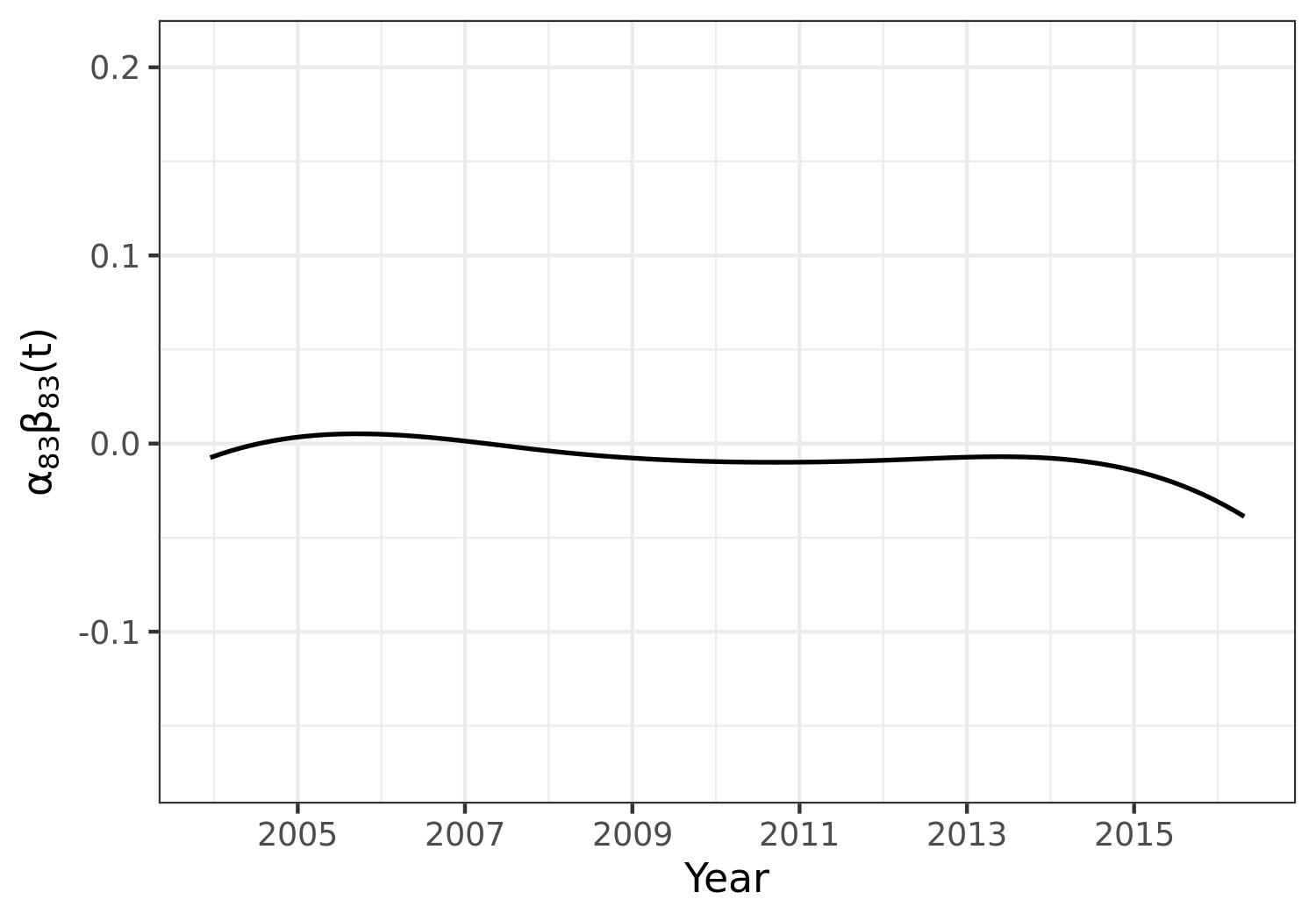}
	\end{minipage}
        }
	\subfigure[Indirect effects (No.96).]{
        \begin{minipage}[t]{0.30\textwidth}
		\includegraphics[width=\textwidth]{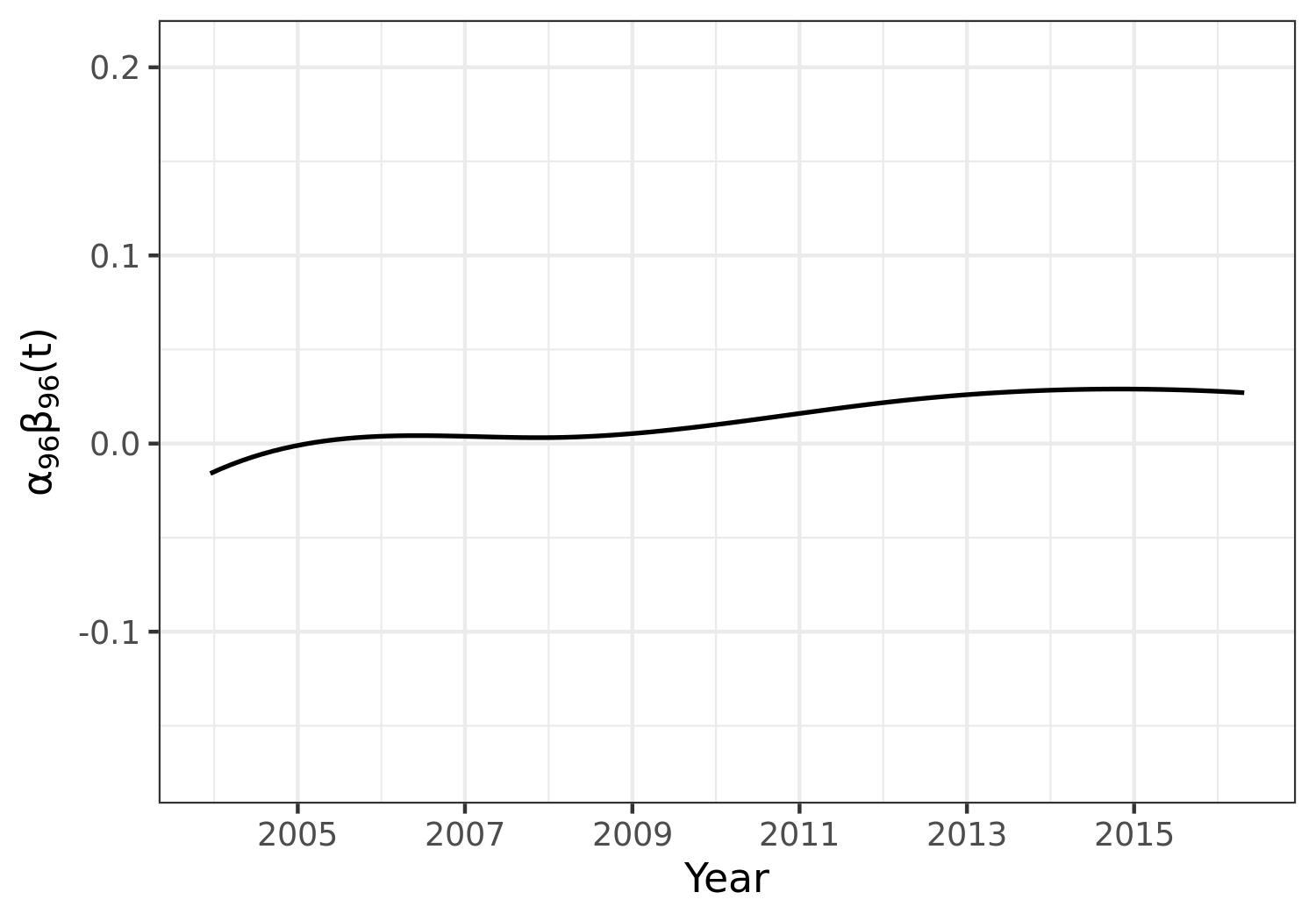}
	\end{minipage}
        }
        \subfigure[Indirect effects (No.438).]{
        \begin{minipage}[t]{0.30\textwidth}
		\includegraphics[width=\textwidth]{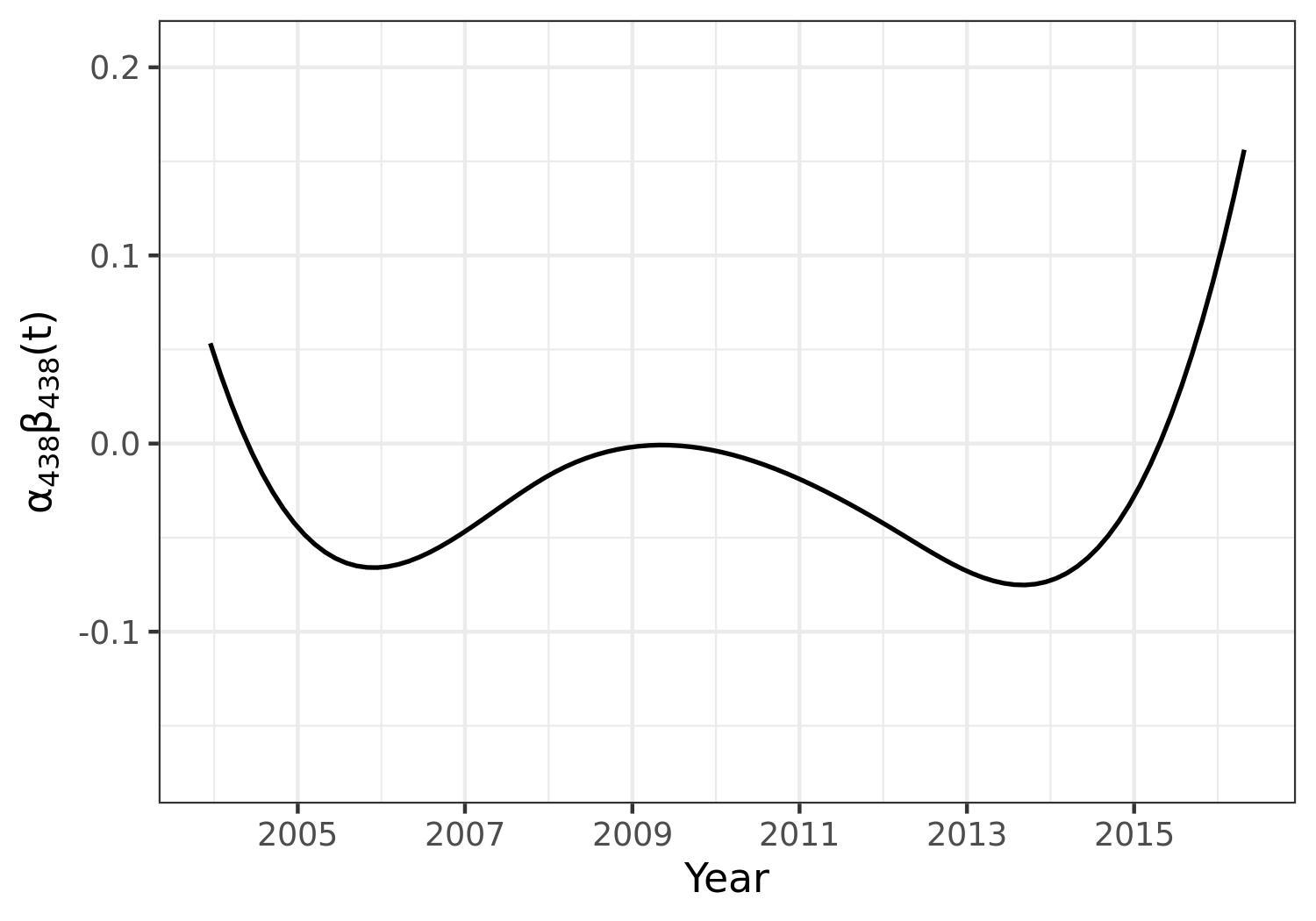}
	\end{minipage}
        }
	\subfigure[Indirect effects (No.440).]{
        \begin{minipage}[t]{0.30\textwidth}
		\includegraphics[width=\textwidth]{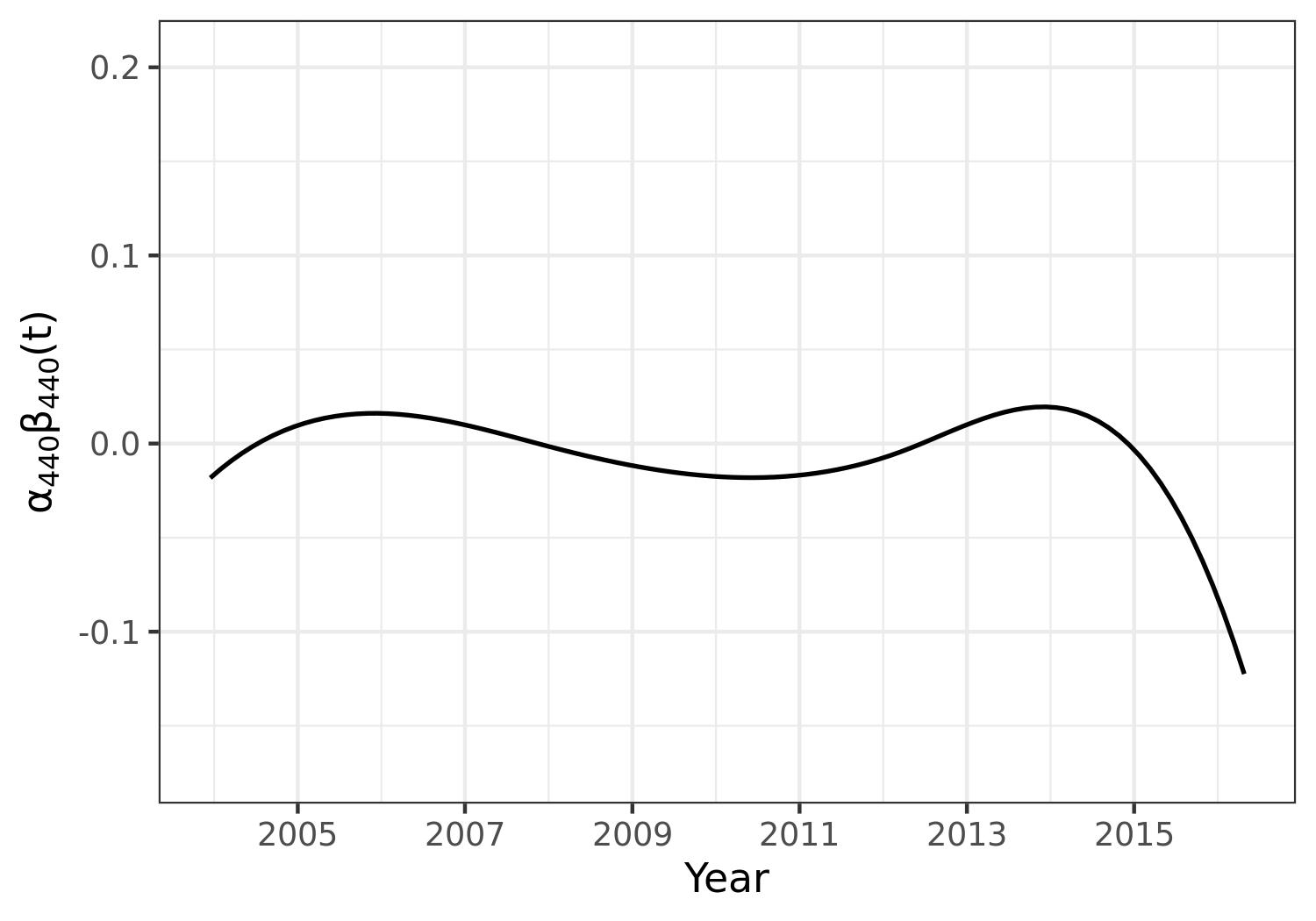}
	\end{minipage}
        }
        \subfigure[Indirect effects (No.472).]{
        \begin{minipage}[t]{0.30\textwidth}
		\includegraphics[width=\textwidth]{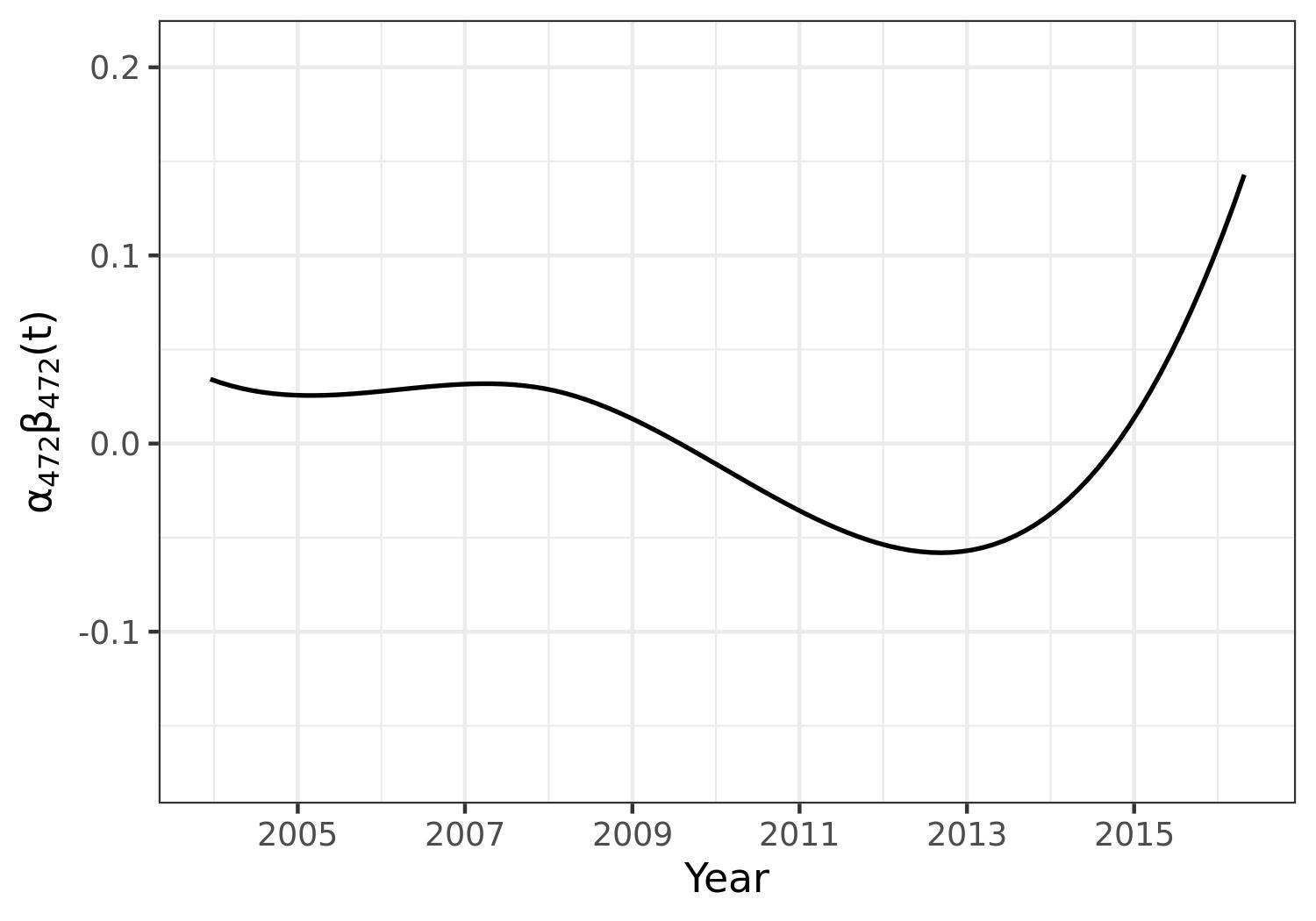}
	\end{minipage}
        }
	\caption{Plots of estimated time-varying direct effect $\eta(t)$ and indirect effect $\alpha_{k}\beta_{k}(t)$ relating to corresponding mediators. }
        \label{fig: effect function}
\end{figure}

We applied our proposed method and the FoSR method by \citet{reiss2011extracting} to identify mediators that exhibit an indirect effect, where $\alpha_{k} \beta_{k}(t) > 0$, while controlling the false discovery rate (FDR) at the 0.05 and 0.1 levels. The comparison method failed to identify any significant mediators at either threshold. Table \ref{tab: Summary results for selected mediators} summarizes the results for each variation of our proposed method based on different user-specified correlation structures. At the 0.1 FDR level, the results showed slight variations depending on the correlation structure, while at the 0.05 FDR level, five common lipid metabolites were consistently identified as significant mediators. These metabolites—{\tt No.83}, {\tt No.96}, {\tt No.438}, {\tt No.440}, and {\tt No.472}—were found to mediate the relationship between acrophase and cognitive decline. Further details of these mediators are summarized in Table \ref{tab: selected mediators}. Particularly, 3-Hydroxytetradecanoyl carnitine ({\tt No.472}) is an acylcarnitine whose general function is to transport chemicals into the mitochondria to produce energy. This is an active research area and open to possibilities of biomarkers \citep{dambrova2022acylcarnitines}. For the specific 3-Hydroxytetradecanoyl carnitine, it is shown altered level in patients with cardiovascular disease and type 2 diabetes mellitus \citep{zhao2020association}.  The five identified mediators and the covariates ($Z$-variables) were included in Models (\ref{M_X}) and (\ref{Y_X+sumM}) to estimate the time-varying mediation effects. Notably, the correlations in the direct effect $\eta(t)$ between $X$ and $Y(t)$ were stronger than the indirect effects $\alpha_{k} \beta_{k}(t)$ between $X$ and $Y$ through $M_{k}$, as shown in Figure \ref{fig: effect function}.

\begin{table}[]
  \centering
  \scriptsize
  \setlength{\tabcolsep}{2pt}
  \caption{Detailed information for consistently selected mediators.}
    \begin{tabular}{ccccc}
    \hline
    Number & SUB PATHWAY & CHEMICAL NAME & HMDB  & PUBCHEM \\
    \hline
    83    & Fatty Acid, Monohydroxy & 3-hydroxymyristate & HMDB0061656 & 16064 \\
    96    & Fatty Acid, Monohydroxy & 3-hydroxylaurate & HMDB0000387 & 94216 \\
    438   & Long Chain Polyunsaturated Fatty Acid (n3 and n6) & tetradecadienoate (14:2)* & HMDB0000560 & 5312409 \\
    440   & Fatty Acid, Dicarboxylate & dodecadienoate (12:2)* & --    & 25480 \\
    472   & Fatty Acid Metabolism (Acyl Carnitine, Hydroxy) & 3-hydroxyoctanoyl-carnitine (1) & HMDB0061634 & 86583357 \\
    \hline
    \end{tabular}
  \label{tab: selected mediators}
\end{table}

The five identified lipid markers in our study were all fatty acids or their carnitine conjugates (i.e., acylcarnitines). Previous studies have shown that altered sleep timing, as measured by the midpoint of sleep, was associated with widespread changes in fatty acids and acylcarnitines. It is noteworthy that two of the metabolites, 3-hydroxymyristate and 3-hydroxylaurate, are metabolites of two saturated long-chain fatty acids, myristic acid and lauric acid. Previous studies have found that elevated levels of these fatty acids are associated with increased risk of dementia. Therefore, it is plausible that alterations in fatty acid metabolism pathways and these metabolites, specifically at least partially, mediate the effect of impaired rest-activity rhythms on cognitive decline. Future studies are needed to elucidate their role in dementia risk among older adults with impaired circadian function.

\section{Discussion}
\label{sec: Discussion}
We proposed a multiple-testing procedure specifically tailored to detect high-dimensional mediation effects in the presence of sparse and irregularly spaced longitudinal outcomes. A functional-regression-based mediation analysis framework was developed to effectively incorporate the within-subject correlation structure across repeated measures in this paper. Additionally, we implemented a resampling-based permutation test to complete the ``JS-mixture'' multiple-testing procedure for identifying significant mediators. The proposed permutation-based hypothesis testing procedure can also be directly applied in time-varying coefficient regression models outside the mediation analysis framework. To address methodological challenges in modeling within-subject correlation, we considered various correlation structures. The consistent selection of mediators in both simulation studies and a real application demonstrates the efficiency and robustness of the proposed method, even when the assumed correlation structure does not match the true underlying structure.

%The utility of our approach was demonstrated through its application to the MrOS Sleep cohort, where we identified lipid pathways potentially mediating the association between rest-activity rhythms and cognitive decline in older adults. The identification of causal effects in our framework relies on a set of structural assumptions. In particular, the assumptions of no unmeasured confounding—essential for identifying natural direct and indirect effects—are unverifiable in empirical studies. In our application we have adjusted for all major confounders based on substantive biological knowledge. However, as with most observational mediation analyses, the presence of unmeasured confounding cannot be entirely ruled out. Designing a sensitivity analysis in complex settings such as ours is nontrivial, as it typically requires additional, untestable structural and modeling assumptions \citep{imai2010identification}. This likely explains why sensitivity analyses are rarely conducted in the causal mediation literature, despite their clear value \citep{zeng2021causal}. We believe it is important to acknowledge this limitation in applied studies explicitly, interpret findings with appropriate caution, and, where feasible, design and conduct sensitivity analyses in future research.

\bibliographystyle{agsm.bst}
\bibliography{bibfile}
\end{document}